\documentclass[%
 aip,
 amsmath,amssymb,
 reprint,%
]{revtex4-1}

\usepackage{graphicx}
\usepackage{dcolumn}
\usepackage{bm}

\usepackage[utf8]{inputenc}
\usepackage[T1]{fontenc}
\usepackage{mathptmx}
\usepackage{etoolbox}

\usepackage{xcolor}
\usepackage[normalem]{ulem}
\makeatletter
\def\@email#1#2{%
 \endgroup
 \patchcmd{\titleblock@produce}
  {\frontmatter@RRAPformat}
  {\frontmatter@RRAPformat{\produce@RRAP{*#1\href{mailto:#2}{#2}}}\frontmatter@RRAPformat}
  {}{}
}%
\makeatother
\begin{document}

\preprint{AIP/123-QED}

\title{Optical decoherence in Er$^{3+}$-doped CeO$_2$ spin qubit platforms}
\author{Vrindaa Somjit*}
\affiliation{Materials Science Division, Argonne National Laboratory, Lemont, Illinois 60439, USA}
\author{Ignas Masiulionis}
\affiliation{Pritzker School of Molecular Engineering, The University of Chicago, Chicago, Illinois 60637, USA}
\affiliation{Materials Science Division, Argonne National Laboratory, Lemont, Illinois 60439, USA}
\author{Gregory D. Grant}
\affiliation{Materials Science Division, Argonne National Laboratory, Lemont, Illinois 60439, USA}
\affiliation{Pritzker School of Molecular Engineering, The University of Chicago, Chicago, Illinois 60637, USA}
\author{Weiguo Jing}
\affiliation{Institute of Condensed Matter and Nanosciences, Université catholique de Louvain,
B-1348 Louvain-la-Neuve, Belgium}
\author{Matteo Giantomassi}
\affiliation{Institute of Condensed Matter and Nanosciences, Université catholique de Louvain,
B-1348 Louvain-la-Neuve, Belgium}
\author{Supratik Guha}
\affiliation{Materials Science Division, Argonne National Laboratory, Lemont, Illinois 60439, USA}
\affiliation{Q-NEXT, Argonne National Laboratory, Lemont, Illinois 60439, USA}
\affiliation{Pritzker School of Molecular Engineering, The University of Chicago, Chicago, Illinois 60637, USA}
\author{Gian-Marco Rignanese}
\affiliation{Institute of Condensed Matter and Nanosciences, Université catholique de Louvain,
B-1348 Louvain-la-Neuve, Belgium}
\author{F. Joseph Heremans}
\affiliation{Materials Science Division, Argonne National Laboratory, Lemont, Illinois 60439, USA}
\affiliation{Q-NEXT, Argonne National Laboratory, Lemont, Illinois 60439, USA}
\affiliation{Pritzker School of Molecular Engineering, The University of Chicago, Chicago, Illinois 60637, USA}
\author{Jiefei Zhang}
\affiliation{Materials Science Division, Argonne National Laboratory, Lemont, Illinois 60439, USA}
\affiliation{Q-NEXT, Argonne National Laboratory, Lemont, Illinois 60439, USA}
\affiliation{Pritzker School of Molecular Engineering, The University of Chicago, Chicago, Illinois 60637, USA}
\author{Giulia Galli*}
\affiliation{Pritzker School of Molecular Engineering, The University of Chicago, Chicago, Illinois 60637, USA}
\affiliation{Department of Chemistry, The University of Chicago, Chicago, Illinois 60637, USA}
\affiliation{Materials Science Division, Argonne National Laboratory, Lemont, Illinois 60439, USA}
\email{vsomjit@anl.gov; gagalli@uchicago.edu}

\date{\today}

\begin{abstract}
Erbium ions (Er$^{3+}$) in cerium dioxide (CeO$_2$) represent a promising spin-photon interface for quantum communication, but the mechanisms limiting their optical coherence remain poorly understood. Using periodic hybrid density functional theory calculations with finite-size corrections, we identify Ce$^{3+}$ polarons and their complexes with oxygen vacancies and Er$^{3+}$ dopants as likely sources of optical decoherence. These defects exhibit finite photoionization cross-sections at 0.8 eV,  coinciding with both the laser excitation energy used experimentally and the emission energy of Er$^{3+}$. This resonance enables photoionization of the polarons and photoluminescence quenching of Er$^{3+}$, leading to the broadening of optical linewidths, shortening of excited-state lifetimes, and introduction of charge noise. Our concentration-dependent photocurrent measurements in Er$^{3+}$-doped CeO$_2$ films under 0.8 eV illumination validate the predicted decoherence pathway. Our combined computational and experimental results identify a concrete defect-engineering target for improving the Er$^{3+}$-doped CeO$_2$ platform, and point to a decoherence mechanism likely relevant to other Er$^{3+}$-doped multivalent-oxide quantum platforms. 
\end{abstract}

\maketitle


Cerium dioxide (CeO$_2$, ceria) is a prototypical mixed ionic-electronic conductor. Due to its mixed valence nature, where Ce$^{4+}$ can be reduced to Ce$^{3+}$ to form electron polarons, and its facile oxygen vacancy formation, ceria is a good candidate for electrochemical and thermochemical energy conversion applications \cite{montini2016fundamentals, li2021review, sun2012nanostructured}. These  include its use for thermochemical solar fuel production \cite{chueh2010high}, as a catalyst in automotive exhaust systems \cite{trovarelli1996catalytic}, and as an electrolyte \cite{steele2001materials} and anode \cite{atkinson2004advanced} in fuel cells. The technological importance of CeO$_2$ has resulted in extensive investigations of its ionic and electronic conductivity mediated by oxygen non-stoichiometry and polaronic transport \cite{tuller1977small}, and of its properties as a function of dopant concentration, dislocations, grain boundaries, heterointerfaces, and strain \cite{andersson2006optimization, sun2015edge, guo2006electrical, lee2020colossal,souza2012modifying}.

Recently, CeO$_2$ has been predicted to be a promising host for spin qubits for quantum technologies \cite{kanai2022generalized}. The low concentration of nuclear spin isotopes (none in Ce and 0.04 \% in O) is expected to lead to a long electron spin coherence time, predicted to be 47 ms in the regime of nuclear spin-limited noise \cite{kanai2022generalized}. CeO$_2$ is also advantageous due to its small lattice mismatch with silicon ($-$0.4 \%, orientation-independent) \cite{tye1994electrical}, facilitating scalability and integration with quantum devices. 

Er$^{3+}$ is a promising spin defect in CeO$_2$, particularly as a spin-photon interface for quantum memories, due to its effective $S=1/2$ ground state, sharp optical transitions in the telecom-C band, and long coherence times observed in several hosts \cite{gupta2025dual, berkman2025long, asadi2018quantum, le2021twenty, gupta2025erbium}. Recently, several studies have investigated Er$^{3+}$-doped CeO$_2$ nanocrystals \cite{wong2024coherent} and films \cite{grant2024optical, zhang2024optical, seth2025spin} as spin qubit platforms. Characterization of films on Si has revealed that Er$^{3+}$ substitutes a Ce$^{4+}$ site and has several favorable properties for use as a quantum memory; these include an optical coherence time of 0.72 $\mu$s, optical excited state lifetime of 3.5 ms, and Hahn-echo spin coherence time ($T_{2,\mathrm{spin}}$) of 0.25 $\mu$s \cite{grant2024optical, zhang2024optical} at 4 K. Encouragingly, $T_{2,\mathrm{spin}}$ has been increased to 40 $\mu$s at 77 mK and found to be limited by spectral diffusion due to surrounding Er$^{3+}$ electron spins \cite{seth2025spin}. 

Although spin decoherence in Er$^{3+}$-doped CeO$_2$ films is now partially understood \cite{seth2025spin}, its optical decoherence mechanism remains unclear. For example, its optical homogeneous linewidth is $\sim 10^3 \times$ higher than its lifetime-limited linewidth, the reason for which is uncertain \cite{grant2024optical, zhang2024optical}. Additionally, electron paramagnetic resonance spectra show a resonant peak with $g=2.04$, whose origin has not yet been fully clarified \cite{zhang2024optical}. 

Microstructural characterization of Er$^{3+}$-doped CeO$_2$ films on Si shows a mixed CeO$_\mathrm{x}$-SiO$_\mathrm{y}$ interfacial layer and the presence of threading dislocations \cite{grant2024optical}. These findings, together with the observed trends in several optical properties as a function of Er$^{3+}$ concentration and annealing temperature \cite{grant2024optical, zhang2024optical}, suggest that optical decoherence could be caused by charged defects, electric field gradients, and strain. In fact, it is reasonable to expect that the very defects beneficial for energy conversion applications (oxygen vacancies and polarons) may be detrimental for quantum ones. Hence, understanding the factors that affect the optical properties of Er$^{3+}$-doped CeO$_2$ films is an important step in the realization of the compelling advantages of both CeO$_2$ and Er$^{3+}$ for quantum applications. 

In this work, we investigate the impact of oxygen vacancies and polarons on the optical decoherence of Er$^{3+}$-doped CeO$_2$ using first-principles calculations.  We predict that the thermodynamic charge transition levels of the isolated polaron and its complexes are below 0.8 eV (1530 nm), the wavelength at which optical transitions in Er$^{3+}$ occur. Importantly, we establish that these defects have a finite photoionization cross-sections at $\sim$0.8 eV. These results indicate that the isolated polaron and its complexes can cause photoluminescence quenching of Er$^{3+}$ emission and can ionize under the 0.8 eV laser illumination used for Er$^{3+}$ excitation. This could lead to optical decoherence by increasing linewidths, reducing optical excited state lifetimes, and causing charge noise. We validate our computational findings through concentration-dependent photocurrent measurements in Er$^{3+}$-doped CeO$_2$ under 0.8 eV laser illumination.

Density functional theory (DFT) calculations were carried out with Quantum Espresso \cite{giannozzi2009quantum, giannozzi2017advanced, giannozzi2020quantum}. We used ONCV pseudopotentials \cite{hamann2013optimized, bosoni2024verify} and the dielectric-dependent hybrid (DDH) exchange-correlation functional (with Hartree-Fock mixing parameter $\alpha$ = $\varepsilon_\infty^{-1}$, where $\varepsilon_\infty=5.31$, as experimentally measured for CeO$_2$ \cite{mochizuki1982infrared})). Our computational setup ensured accurate bandgap predictions and the localization of the Ce$^{3+}$ polaron. The computed optical bandgap (with TDDFT@DDH) is 3.49 eV, consistent with the value of $\sim$3.55 eV obtained from UV-VIS and reflectivity measurements \cite{zhang2006optical, marabelli1987covalent, guo1995spectroscopic} and from our photoluminescence emission and excitation measurements (see Supplementary Material SM Sections I--III).

We computed the thermodynamic charge transition levels (CTL) and optical charge transition levels (or vertical transition levels, VTL). The former corresponds to the thermal ionization energy, in which an electron (hole) is ionized into the CBM (VBM), thereby changing the charge state of the defect from charge $q$ to $q'$, after which the defect relaxes to the equilibrium configuration corresponding to $q'$. The CTL (from the VBM) is computed as \cite{freysoldt2014first}
\begin{equation}
    \label{eq:td_ctl}
    \varepsilon(q / q') = \frac{E_\mathrm{f}\left(q, \mathbf{R}_{q}; \varepsilon_{\mathrm{F}} = 0\right) - E_\mathrm{f}\left(q', \mathbf{R}_{q'}; \varepsilon_{\mathrm{F}} = 0\right)}{q' - q},
\end{equation}
and is the central quantity required to obtain the photoionization cross-section; $E_\mathrm{f}(q, \mathbf{R}_{q'})$ is the formation energy of a defect in charge state $q$ and configuration $\mathbf{R}_{q'}$ \cite{freysoldt2014first, falletta2020finite} (see SM Section IV for calculation of $E_\mathrm{f}(q, \mathbf{R}_{q'})$). 

The VTL corresponds to the optical ionization energy, in which an electron (hole) is ionized into the CBM (VBM), thereby changing the charge state of the defect from $q$ to $q'$; the defect remains frozen in the equilibrium configuration corresponding to $q$ over the timescale of the measurement. The VTL (from the VBM) is computed as~\cite{freysoldt2014first, gake2020finite, falletta2020finite}
\begin{equation}
    \label{eq:vtl}
    \mu(q / q', \mathbf{R}_{q}) = \frac{E_\mathrm{f}\left(q, \mathbf{R}_{q}; \varepsilon_{\mathrm{F}} = 0\right) - E_\mathrm{f}\left(q', \mathbf{R}_{q}; \varepsilon_{\mathrm{F}} = 0\right)}{q' - q},
\end{equation}
and is the central quantity required to understand absorption and emission processes involving carrier exchange with the host material. 

In periodic calculations, computing accurate values of CTLs and VTLs requires the evaluation of correction terms $E_{\mathrm{corr}}$ that account for the artificial electrostatic interactions arising due to the finite supercell size. We use the expression proposed in Ref.~\cite{freysoldt2009fully} to compute $E_{\mathrm{corr}}\left(q, \mathbf{R}_{q}\right)$  (with $\varepsilon_0=24.5$ \cite{mochizuki1982infrared} for CeO$_2$) and that of Ref.~\cite{gake2020finite} to compute $E_{\mathrm{corr}}\left(q, \mathbf{R}_{q'}\right)$ (with $\varepsilon_{\infty}=5.31$ \cite{mochizuki1982infrared} for CeO$_2$), as implemented in sxdefectalign \cite{sxdefectalign}. The evaluation of $E_{\mathrm{corr}}\left(q, \mathbf{R}_{q'}\right)$ requires special care, as different ionic and electronic screening terms are involved when considering a charge $q$ in a fixed configuration corresponding to that of the charge $q'$ \cite{gake2020finite, falletta2020finite}.

Finally, the photoionization cross-section is the transition rate per unit photon flux and, for an ensemble of randomly oriented defects, is computed as \cite{stoneham2001theory, razinkovas2021photoionization}
\begin{equation}
    \sigma_{\text{ph}}(\hbar\omega) = \frac{4\pi^{2}\alpha}{3 n_{\mathrm{ref}}} \hbar\omega \sum_j |\mathbf{r}_{dj}|^{2} A(\hbar\omega - E_{dj}),
    \label{eq:photoion_cs}
\end{equation}

where $\alpha$ is the fine-structure constant, $n_{\mathrm{ref}}$ is the refractive index of CeO$_2$ (2.30), $\mathbf{r}_{dj}$ is the optical matrix element between the Kohn-Sham states $\varphi_d$ and $\varphi_j$, computed using  the WEST code \cite{govoni2015large, yu2022gpu, marcks2024quantum}, $\hbar\omega$ is the photon energy. The sum runs over all final Kohn-Sham states $\varphi_j$ (the entire Ce 4f band);  $E_{dj}$ is the energy difference between the Kohn-Sham eigenvalues of the initial (defect Kohn-Sham state, $\varphi_d$) and final states, rigidly shifted so that the smallest value of $E_{dj}$ corresponds to the CTL. $A(\hbar\omega - E_{dj})$ is the normalized spectral function of electron-phonon coupling that accounts for the vibrational overlap between the ground and excited states and is computed using the one-dimensional configurational coordinate diagram (1D-CCD) approach (see SM Section V).

\begin{figure}[t] 
    \centering
    \includegraphics[width=0.88\linewidth]{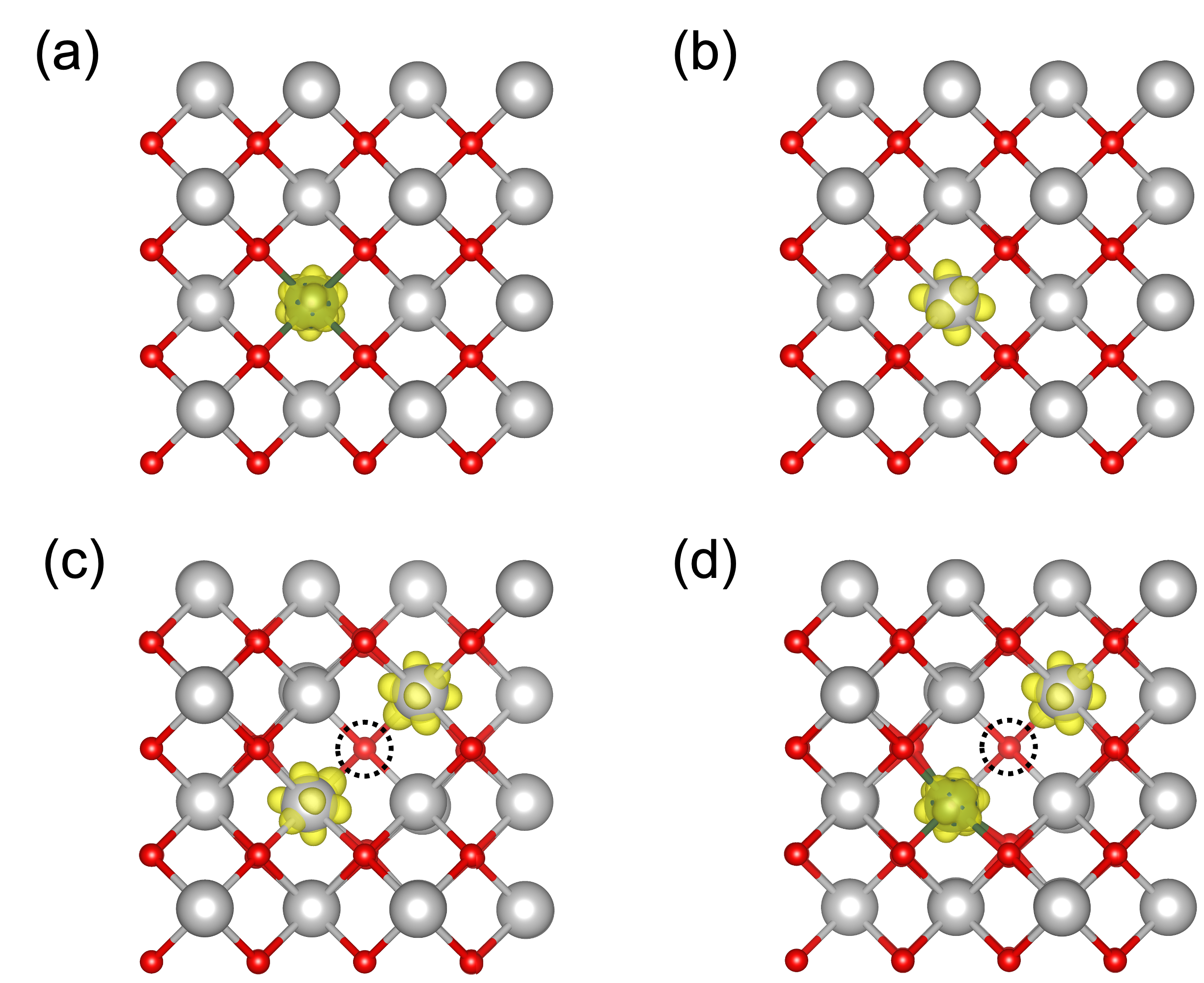}
    \caption{Isosurfaces of spin polarization densities of several defects investigated in this work. (a) $\mathrm{Er_{Ce}^{1-}}$ (b) $\mathrm{Ce_{Ce}^{1-}}$ (c) $\mathrm{V_O^0}$ (i.e., $\mathrm{[Ce_{Ce}^{1-}-V_O^{2+}-Ce_{Ce}^{1-}]^0}$) (d) $\mathrm{[Er_{Ce}-V_O]^0}$ (i.e., $\mathrm{[Er_{Ce}^{1-}-V_O^{2+}-Ce_{Ce}^{1-}]^0}$). Defects are described using the expression $\mathrm{A_B}^q$, denoting atom A substituting site B with relative charge $q$. $\mathrm{V_O}$ denotes oxygen vacancy.
    Cerium, oxygen, and erbium are represented by gray, red and green spheres, respectively and vacancy by dotted circle. Isosurface level is 8.5 $\times$ 10$^{-3}$ e$^-$/\AA$^3$.}
    \label{fig:isosurfaces}
\end{figure}

Figure \ref{fig:isosurfaces} shows the isosurfaces of the spin polarization densities of several defects considered in our study. We investigated the Er$^{3+}$ substitutional dopant on the Ce$^{4+}$ site ($\mathrm{Er_{Ce}}$) in the $-1$ and 0 charge states. As seen in Figure \ref{fig:isosurfaces}a, our calculations correctly reproduce the localization of the Er 4f orbitals. We also investigated the isolated electron polaron ($\mathrm{Ce_{Ce}^{1-}}$, Figure \ref{fig:isosurfaces}b); oxygen vacancy ($\mathrm{V_O^q}$) in the 0, +1, and +2 charge states; and oxygen vacancy complexes with substitutional Er$^{3+}$ ($\mathrm{[Er_{Ce}-V_O]^q}$) in the 0 and +1 charge states. We note that all the defects investigated here (except the isolated Er dopant and isolated polaron) are complexes with a polaron (see  Figure \ref{fig:isosurfaces}c-d).  

\begin{table}[t] 
\caption{\label{tab:CTL_VTL} Energy (eV) of the thermodynamic charge transition levels (CTLs, Equation \ref{eq:td_ctl}) and vertical transition levels (VTLs, Equation \ref{eq:vtl}) relative to the CBM (minimum of the Ce 4f band) for the defects investigated in this work. The difference between the CTL of the defect complex and that of the isolated polaron is the binding energy of the complex. All  binding energy values are positive, indicating the stability of the complexes. $\mathrm{V_O}$ denotes oxygen vacancy.}
\begin{ruledtabular}
\begin{tabular}{l c c c c}
Defect & CTL & Energy & VTL & Energy \\
\hline
$\mathrm{Er_{Ce}}$     & (0/$-$1)  & 3.11 &                &       \\ [4 pt]
Polaron                & (0/$-$1)  & 0.62 &  ($-$1/0)      & 1.40  \\
                       &           &      &  (0/$-$1)      & $-$0.05 \\ [4 pt]
$\mathrm{V_O}$         & (+2/+1) & 0.74 &  (+1/+2)     & 1.89  \\
                       &           &      &  (+2/+1)     & 0.09  \\ [4 pt]
$\mathrm{V_O}$         & (+1/0)  & 0.81 &  (0/+1)      & 1.56  \\
                       &           &      &  (+1/0)     & 0.22  \\ [4 pt]
$[\mathrm{Er_{Ce}-V_O}]$ & (+1/0)  & 0.84 &  (0/+1)      & 1.91  \\
                       &           &      &  (+1/0)      & 0.23  \\
\end{tabular}
\end{ruledtabular}
\end{table}

Table \ref{tab:CTL_VTL} lists the computed CTLs of the various defects with respect to the CBM. We first discuss the polaron self-trapping energy of the isolated polaron (also known as polaron formation energy or polaron binding energy), which is related to the CTL as $E_{\mathrm{self-trap}}=-\varepsilon(0/-1)_{\mathrm{pol}}$. Our computed value, $-0.62$ eV, lies within the range of values reported in previous \textit{ab initio} works, which vary from $-0.3$ to $-1.8$~eV (see SM Section VII) \cite{castleton2019benchmarking, sun2017disentangling}. This variance in the literature values is due to the functional choice and, importantly, to the lack of appropriate finite-size corrections in many of the computed values of $E_{\mathrm{self-trap}}$. Crucially, as we discuss in the SM, the use of the DDH functional and proper finite-size corrections ensures piecewise linearity of the exchange-correlation functional, confirming the robustness of our results for $E_{\mathrm{self-trap}}$. While our computed $E_{\mathrm{self-trap}}$ is slightly higher than that measured using room temperature transient absorption spectroscopy ($-0.4$ eV) \cite{pelli2020ultrafast}, the agreement with experiment is still satisfactory, since our calculations are at 0~K and do not include renormalization of the bandgap and of $E_\mathrm{self-trap}$ due to finite temperature and nuclear quantum effects.

The CTL of Er$^{3+}$ is over 3 eV below the CBM, indicating a large ionization energy. Thus, photoionization of the Er$^{3+}$ spin defect under operating conditions is an unlikely decoherence mechanism. Interestingly, the CTLs of all the polaron-related defects, either the isolated polaron or its complexes with $\mathrm{V_O^{2+}}$ and $\mathrm{[Er_{Ce}-V_O]^{1+}}$, are $\lesssim$ 0.8~eV. Thus, these defects could potentially photoionize under the 0.8 eV laser illumination used to excite Er$^{3+}$, or cause photoluminescence quenching by absorbing the 0.8 eV emission from Er$^{3+}$. Whether these processes occur is decided by their photoionization cross-sections (discussed below).

Table 1 also lists the VTLs of the various defects with respect to the CBM. Importantly, the $\mu(q/q-1, \mathbf{R}_q)$ VTLs, which reflect the de-excitation of a delocalized electron in the Ce 4f band to form a localized polaron (isolated or associated with a defect), are all relatively low, indicating that the non-radiative carrier capture of the delocalized electron is likely. In particular, $\mu(0/-1)_{\mathrm{pol}}$ is negative ($-$0.05 eV), which implies that the delocalized electron will undergo vibrational relaxation (on a timescale of picoseconds, i.e., THz rates) to form a polaron. Thus, the $g=2.04$ peak seen in EPR measurements \cite{zhang2024optical} is likely due to the Ce$^{3+}$ polaron and not a delocalized electron.

CTLs only indicate which defects may ionize in our energy range of interest; computing both the transition dipole moment and the vibrational overlap between the ground and excited states is necessary to confirm whether the transitions occur. These are accounted for in the photoionization cross-sections. 

\begin{figure}[t] 
    \centering
    \includegraphics[width=1.0\linewidth]{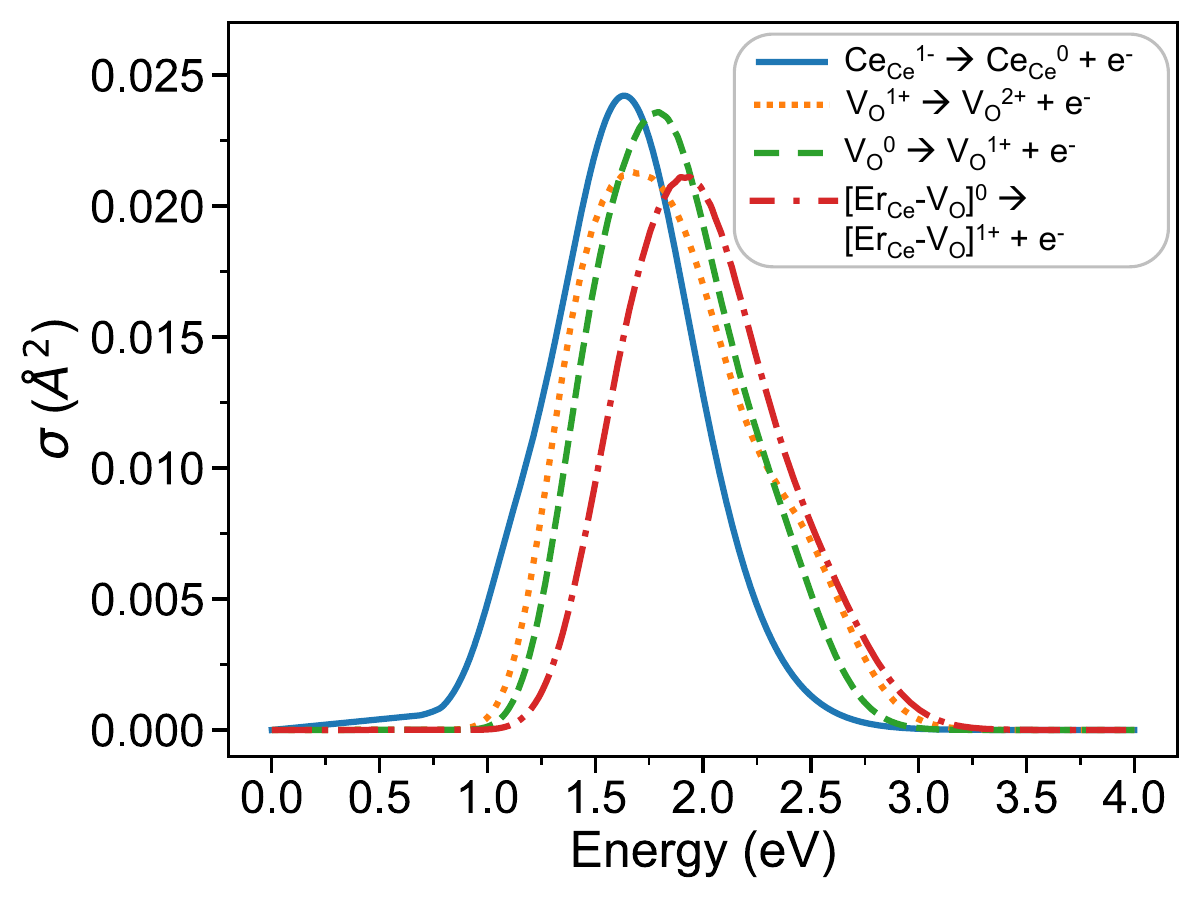}
    \caption{Computed photoionization cross-sections with vibrational broadening at 4 K. Notation in legend described in Figure \ref{fig:isosurfaces}.}
    \label{fig:photoion_cs}
\end{figure}

Figure \ref{fig:photoion_cs} shows the photoionization cross-sections with vibrational broadening (see SM Sections VIII and IX for the cross-sections without vibrational broadening, the 1D-CCD plots, and effective phonon frequencies used to compute $A(\hbar\omega - E_{dj})$). We compute the cross-sections at 4 K to reflect the temperature at which optical and spin characterization measurements are usually performed \cite{grant2024optical, zhang2024optical} (the cross-sections are similar at 300 K, see SM). Notably, the cross-section at 0.8 eV is significant for the ionization of the isolated polaron. Although the photoionization onset is slightly higher than 0.8 eV for the polaron-defect complexes, given the approximate values of the computed CTLs (due to the absence of finite temperature and nuclear quantum effects) and expected bandgap reduction at finite temperatures \cite{han2023temperature}, it is possible that a finite photoionization cross-section occurs at 0.8 eV for the defect complexes as well.

Together, our results suggest the following mechanism for optical decoherence in Er$^{3+}$-doped CeO$_2$. The isolated polaron and its complexes can cause photoluminescence quenching by absorbing the emission from Er$^{3+}$ and can photoionize under the 0.8 eV laser used to excite Er$^{3+}$. Photoluminescence quenching of Er$^{3+}$ can then lead to broadening of optical linewidths and to the reduction of optical excited state lifetimes. Additionally, the impurities ionized by emission from Er$^{3+}$ or by the laser can form another polaron or polaron-defect complex, causing charge noise. 

\begin{figure}[t] 
    \centering
    \includegraphics[width=1.0\linewidth]{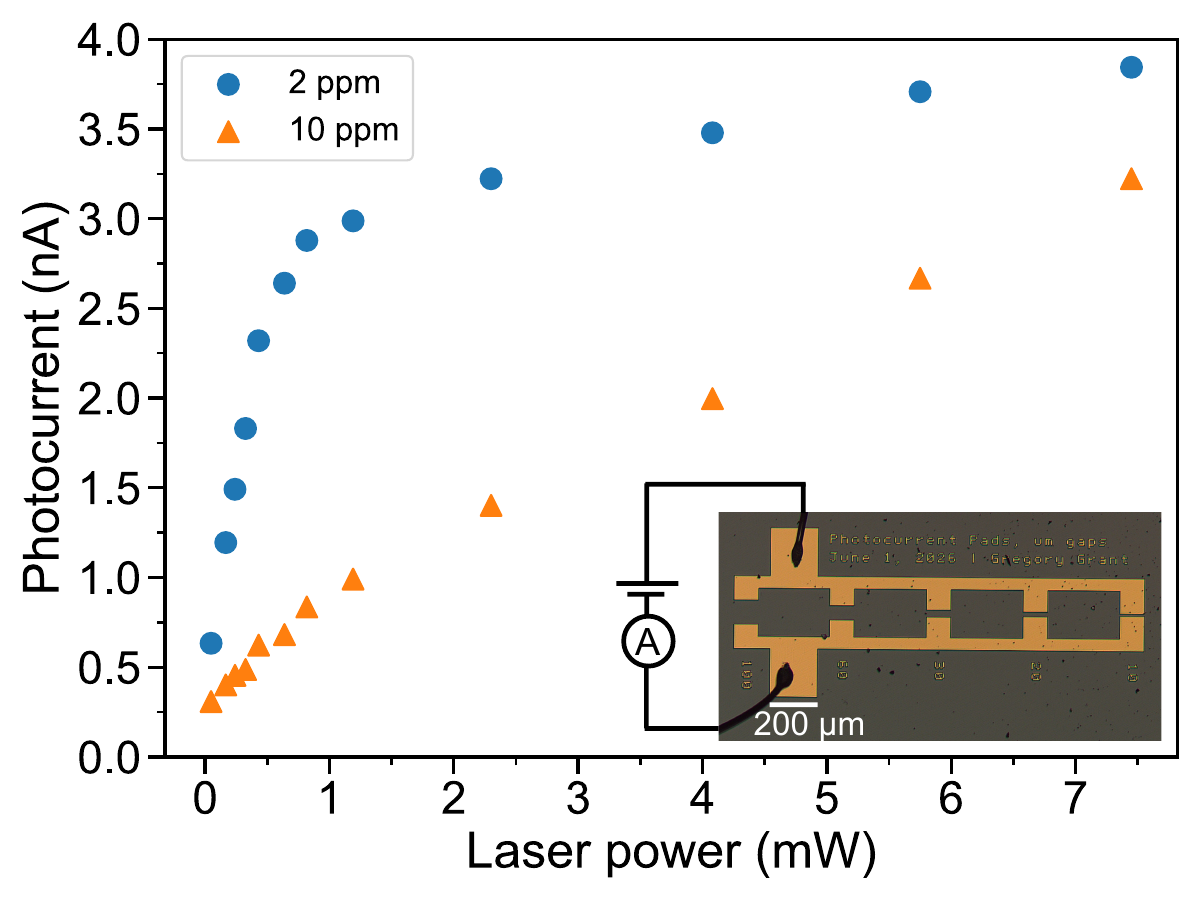}
    \caption{Measured photoexcited current at 300 K as a function of laser power under 5 V gate bias across a 60 $\mu$m gap in Er$^{3+}$-doped CeO$_2$ films with two different Er$^{3+}$ concentrations. Laser focus was kept at 30 $\mu$m out of the sample. Photocurrent is defined as $I_{\mathrm{photo}}=I-I_{\mathrm{dark}}$, where $I$ and $I_{\mathrm{dark}}$ are the total current and current measured with laser off, respectively. Inset: Ti/Au gates (gold) deposited on Er$^{3+}$-doped CeO$_2$ (brown) with varying gaps.} 
    \label{fig:photocurrent}
\end{figure}

To validate the predicted  polaron-mediated optical decoherence pathway, we measured the photocurrent under 0.8 eV laser illumination in two single crystal $\sim$750 nm-thick CeO$_2$ films doped with 2 and 10 ppm Er$^{3+}$. Our samples were grown using molecular beam epitaxy \cite{grant2024optical}. Ti/Au gates were fabricated on the samples using standard photolithography techniques (Figure \ref{fig:photocurrent} inset). Standard I-V measurements established the presence of charge carriers in the samples (see SM Section X), and subsequent photocurrent measurements were carried out by applying a voltage across the two gates under 0.8 eV (1530 nm) laser illumination. The resulting current with increasing laser power was measured with an electrometer.

Figure \ref{fig:photocurrent} shows that both samples give rise to photoexcitation-induced current. We verified that the photocurrent is dominated by carriers from CeO$_2$ bulk (not the surface; see SM); this result confirms the presence of charged defects in Er$^{3+}$-doped CeO$_2$ that photoionize under laser illumination used for Er$^{3+}$ excitation, as predicted by our calculations. These defects likely arise due to the formation of the interfacial CeO$_\mathrm{x}$-SiO$_\mathrm{y}$ layer, which reacts with oxygen from the CeO$_2$ lattice, leaving behind oxygen vacancies compensated by Ce$^{3+}$ polarons. 

\begin{figure}[t] 
     \centering
     \includegraphics[width=1.0\linewidth]{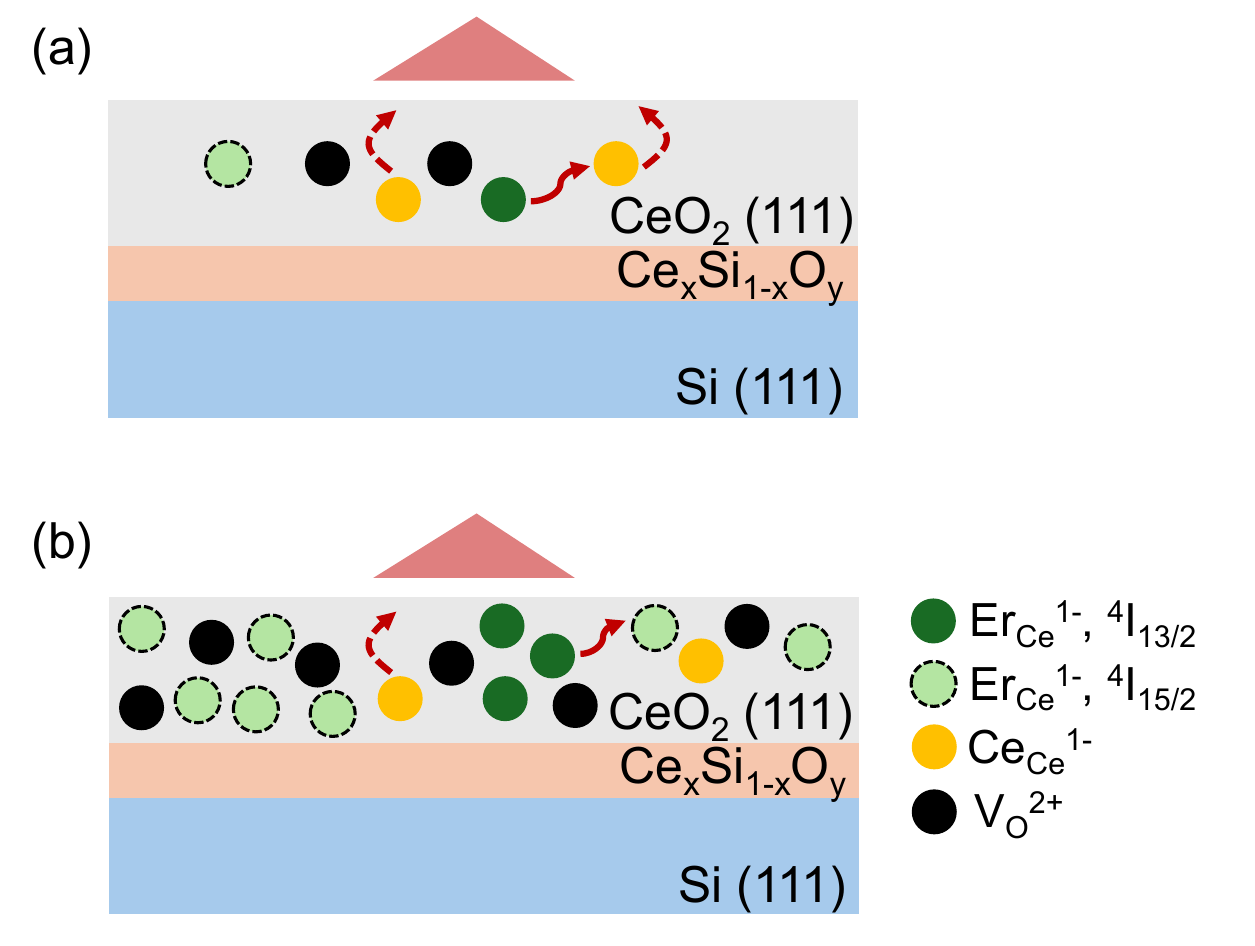}
     \caption{Identified optical decoherence pathway in Er$^{3+}$-doped CeO$_2$ with (a) 2 ppm (b) 10 ppm Er$^{3+}$ (see text). Red triangle denotes laser illumination. Solid arrow denotes energy transfer from excited Er$^{3+}$. Dashed arrow denotes ionization of Ce$^{3+}$. In (a), the large concentration of Ce$^{3+}$ relative to Er$^{3+}$ results in energy transfer from excited Er$^{3+}$ to Ce$^{3+}$, ionizing it, in addition to ionization of Ce$^{3+}$ due to the laser. In (b), the low concentration of Ce$^{3+}$ relative to Er$^{3+}$ results in energy transfer from excited Er$^{3+}$ to ground-state Er$^{3+}$, instead of Ce$^{3+}$. Ionization of Ce$^{3+}$ due to the laser still occurs.}
     \label{fig:schema}
\end{figure}

Interestingly, while the 2 ppm and 10 ppm samples show largely similar I-V curves, indicating similar carrier densities (see SM), they show a significantly different dependence of the photocurrent on laser power. As seen in Figure \ref{fig:photocurrent}, the 2 ppm sample is more photoresponsive compared to the 10 ppm sample and exhibits a saturation effect at high power. 

The distinct concentration-dependent photoresponse can be understood by considering the relative concentration of Ce$^{3+}$ and Er$^{3+}$, as depicted in Figure \ref{fig:schema}. While the \textit{total} concentration of Ce$^{3+}$ is likely the same in both samples (due to the formation of a CeO$_\mathrm{x}$-SiO$_\mathrm{y}$ interfacial layer in both samples), the \textit{relative} concentration of Ce$^{3+}$, compared to Er$^{3+}$, is higher in the 2 ppm sample compared to the 10 ppm sample. Thus, the emitted energy from Er$^{3+}$ is likely transferred more effectively to Ce$^{3+}$ in the 2 ppm sample (Figure \ref{fig:schema}a) and quenched, ionizing Ce$^{3+}$ in the process. In contrast, the higher concentration of Er$^{3+}$ in the 10 ppm sample enhances the probability that Er$^{3+}$ emission is transferred to other unexcited Er$^{3+}$ ions, instead of being transferred to Ce$^{3+}$ (Figure \ref{fig:schema}b). Therefore, the increased probability of energy transfer from Er$^{3+}$ to Ce$^{3+}$ in the 2 ppm sample explains its higher photocurrent. 
The saturation behavior in the 2~ppm sample can also be explained by the difference in the two energy transfer routes: Er$^{3+}$ to Ce$^{3+}$ vs. Er$^{3+}$ to Er$^{3+}$. 
In the 2~ppm sample, the laser and Er$^{3+}$ to Ce$^{3+}$ energy transfer ionize most of the Ce$^{3+}$ ions, so that at higher power, no Ce$^{3+}$  is left to be ionized, thereby saturating the photocurrent. In the 10 ppm sample, the laser ionizes a fraction of the Ce$^{3+}$ ions (similar to the 2 ppm case); however, the more probable Er$^{3+}$ to Er$^{3+}$ energy transfer prevents ionizing all of Ce$^{3+}$. Hence, unionized Ce$^{3+}$ ions remain in the sample, where they can ionize at higher laser powers, preventing saturation. Thus, the observation of a photocurrent confirms the photoionization of charged defects under 0.8 eV laser illumination, and the concentration dependence of the photocurrent confirms that these charged defects can cause photoluminescence quenching of Er$^{3+}$ emission. Together, these processes lead to optical decoherence. 

In summary, we elucidated the role of electron polarons and their complexes with oxygen vacancies and Er$^{3+}$ dopants on the optical decoherence observed in Er$^{3+}$-doped CeO$_2$. The thermodynamic and optical charge transition levels and photoionization cross-sections predicted by hybrid DFT calculations and proper finite-size corrections indicate that polarons and their complexes cause photoluminescence quenching of the Er$^{3+}$ emission and photoionize under the laser used to excite  Er$^{3+}$. These mechanisms cause optical decoherence by increasing linewidths, reducing optical excited state lifetimes, and introducing charge noise. Our concentration-dependent photocurrent measurement further validates the predicted decoherence pathway. The decoherence mechanism proposed here may be relevant for other Er$^{3+}$-doped oxide film platforms that are increasingly being investigated for quantum applications and have mixed-valent cations, such as Er$^{3+}$-doped TiO$_2$ \cite{martins2024role} and CaWO$_4$ \cite{tang2026isotopically} films.  Future work will focus on computing the rates for the various energy transfer processes that occur in  Er$^{3+}$-doped CeO$_2$ and on developing charge noise models. These will further quantify the role of charged defect-mediated decoherence and inform design strategies for its mitigation. 

\section*{Supplementary Material}
Supplementary Material includes details on density functional theory calculations (pseudopotentials, finite-size corrections, piecewise linearity, formation energies, photoionization cross-sections, comparison with previous experimental and computational literature on CeO$_2$) and photocurrent measurements. 

\begin{acknowledgments}
The authors thank Diego Sorbelli, Jiawei Zhan, Yu Jin, and Shreya Verma for helpful discussions. This work was primarily funded by MICCoM, which is part of the Computational Materials Sciences Program funded by the U.S. Department of Energy, Office of Science, Basic Energy Sciences, Materials Sciences, and Engineering Division. V.S acknowledges the support from the Maria Goeppert Mayer Named Fellowship, under the Laboratory Directed Research and Development (LDRD) funding from Argonne National Laboratory, provided by the Director, Office of Science, of the U.S. Department of Energy under Contract No. DE-AC02-06CH11357. The optical and photocurrent measurements were supported by the U.S. Department of Energy, Office of Science, Basic Energy Sciences, Materials Sciences and Engineering Division (F.J.H., J.Z.). The material synthesis and photocurrent device fabrication (I.M., G.D.G, S.G.) were supported by Q-NEXT, a U.S. Department of Energy Office of Science National Quantum Information Science Research Center under Award Number DE-FOA-0002253. G.-M.R. is a Research Director of the Fonds de la Recherche Scientifique - FNRS. W.J. acknowledges the support from FNRS. This research used resources of the National Energy Research Scientific Computing Center (NERSC), a Department of Energy Office of Science User Facility using NERSC award DDR-ERCAP0029604 and resources of the Argonne Leadership Computing Facility, a U.S. Department of Energy (DOE) Office of Science user facility at Argonne National Laboratory, which is supported by the Office of Science of the U.S. DOE under Contract No. DE-AC02-06CH11357.
\end{acknowledgments}

\section*{Data Availability Statement}
The data that support the findings of this study will be made available on Qresp \cite{Qresp}.

\section*{Author Declarations}
The authors have no conflicts to disclose.

\section*{Author contributions}
V.S. - Conceptualization (lead), Data Curation (lead), Formal Analysis (equal), Funding Acquisition (equal), Investigation (equal), Methodology (lead), Project Administration (equal), Resources (equal), Software (equal), Validation (equal), Visualization (lead), Writing/Original Draft Preparation (lead), Writing/Review and Editing (equal) \\
I.M. and G.D.G. - Formal Analysis (equal), Investigation (equal), Validation (equal), Writing/Review and Editing (equal) \\
W.J. and M.G. - Investigation (equal), Methodology (supporting), Software (equal), Validation (equal), Writing/Review and Editing (equal) \\
S.G. - Funding Acquisition (equal), Project Administration (equal), Resources (equal), Supervision (equal), Writing/Review and Editing (equal) \\
G.M.R. - Funding Acquisition (equal), Project Administration (equal), Resources (equal), Supervision (equal), Writing/Review and Editing (equal) \\
F.J.H. - Methodology (supporting), Supervision (equal), Writing/Review and Editing (equal) \\
J.Z. - Methodology (supporting), Project Administration (equal), Resources (equal), Supervision (equal), Writing/Review and Editing (equal) \\
G.G. - Conceptualization (supporting), Project Administration (equal), Resources (equal), Supervision (equal), Writing/Review and Editing (equal)

\bibliography{main}

\end{document}


\preprint{}

\title{Supplementary Material: \\Optical decoherence in Er$^{3+}$-doped CeO$_2$ spin qubit platforms}
\author{Vrindaa Somjit*}
\affiliation{Materials Science Division, Argonne National Laboratory, Lemont, Illinois 60439, USA}
\author{Ignas Masiulionis}
\affiliation{Pritzker School of Molecular Engineering, The University of Chicago, Chicago, Illinois 60637, USA}
\affiliation{Materials Science Division, Argonne National Laboratory, Lemont, Illinois 60439, USA}
\author{Gregory D. Grant}
\affiliation{Materials Science Division, Argonne National Laboratory, Lemont, Illinois 60439, USA}
\affiliation{Pritzker School of Molecular Engineering, The University of Chicago, Chicago, Illinois 60637, USA}
\author{Weiguo Jing}
\affiliation{Institute of Condensed Matter and Nanosciences, Université catholique de Louvain,
B-1348 Louvain-la-Neuve, Belgium}
\author{Matteo Giantomassi}
\affiliation{Institute of Condensed Matter and Nanosciences, Université catholique de Louvain,
B-1348 Louvain-la-Neuve, Belgium}
\author{Supratik Guha}
\affiliation{Materials Science Division, Argonne National Laboratory, Lemont, Illinois 60439, USA}
\affiliation{Q-NEXT, Argonne National Laboratory, Lemont, Illinois 60439, USA}
\affiliation{Pritzker School of Molecular Engineering, The University of Chicago, Chicago, Illinois 60637, USA}
\author{Gian-Marco Rignanese}
\affiliation{Institute of Condensed Matter and Nanosciences, Université catholique de Louvain,
B-1348 Louvain-la-Neuve, Belgium}
\author{F. Joseph Heremans}
\affiliation{Materials Science Division, Argonne National Laboratory, Lemont, Illinois 60439, USA}
\affiliation{Q-NEXT, Argonne National Laboratory, Lemont, Illinois 60439, USA}
\affiliation{Pritzker School of Molecular Engineering, The University of Chicago, Chicago, Illinois 60637, USA}
\author{Jiefei Zhang}
\affiliation{Materials Science Division, Argonne National Laboratory, Lemont, Illinois 60439, USA}
\affiliation{Q-NEXT, Argonne National Laboratory, Lemont, Illinois 60439, USA}
\affiliation{Pritzker School of Molecular Engineering, The University of Chicago, Chicago, Illinois 60637, USA}
\author{Giulia Galli*}
\affiliation{Pritzker School of Molecular Engineering, The University of Chicago, Chicago, Illinois 60637, USA}
\affiliation{Department of Chemistry, The University of Chicago, Chicago, Illinois 60637, USA}
\affiliation{Materials Science Division, Argonne National Laboratory, Lemont, Illinois 60439, USA}
\email{vsomjit@anl.gov; gagalli@uchicago.edu}


\date{\today}

\maketitle

\section{Density functional theory calculations}
Density functional theory (DFT) calculations were carried out within the generalized Kohn-Sham framework \cite{hohenberg1964inhomogeneous, kohn1965self}, using the plane-wave pseudopotential method as implemented in the Quantum Espresso code \cite{giannozzi2009quantum, giannozzi2017advanced, giannozzi2020quantum}. ONCV pseudopotentials \cite{hamann2013optimized, bosoni2024verify} were used, with $\mathrm{5s^25p^66s^24f^15d^1}$, $\mathrm{5s^25p^66s^24f^{11}5d^1}$, and $\mathrm{2s^22p^4}$ as the valence configurations for Ce, Er, and O, respectively. The inclusion of 4f electrons in the valence partition necessitates a large plane-wave cutoff of 200 Ry. The dielectric-dependent hybrid (DDH) exchange-correlation functional was used, with the Hartree-Fock mixing parameter $\alpha$ set to 0.188 ($\alpha$ = $\varepsilon_\infty^{-1}$, where $\varepsilon_\infty=5.31$, as experimentally measured for bulk single crystal CeO$_2$ \cite{mochizuki1982infrared}). Spin-polarized calculations were carried out with a 2 $\times$ 2 $\times$ 2 (96 atoms) supercell of cubic fluorite CeO$_2$, using the experimental lattice constants at 298 K ($a=b=c=5.411$ \AA \cite{swanson1953standard}). The Brillouin zone of the supercell was sampled with the $\Gamma$ point. 

\newpage

\section{Performance of scalar relativistic ONCV pseudopotentials}

Tables \ref{tab:Ce_PP_nu} and \ref{tab:Er_PP_nu} summarize the performance of the Ce and Er ONCV pseudopotentials \cite{hamann2013optimized, bosoni2024verify} developed in this work. Their accuracy was assessed using the $\nu$ metric, which quantitatively measures the difference between the equations of state (EOS) obtained with a pseudopotential (PS) and those from all-electron (AE) calculations. A smaller $\nu$ value indicates better agreement with the AE reference and therefore higher pseudopotential accuracy.

Following the framework proposed by Bosoni et al. \cite{bosoni2024verify},

\begin{equation}
    \nu(PS,AE) = 100 \sqrt{ \sum_{Y = V_0, B_0, B_1}\left[ w_Y \, \frac{Y_{PS} - Y_{AE}}{(Y_{PS} + Y_{AE})/2} \right]^2 },
    \label{eq:nu metric}
\end{equation}

where the benchmark consists of four mono-elemental cubic crystals and six cubic oxides, as summarized in Tables \ref{tab:Ce_PP_nu} and \ref{tab:Er_PP_nu}. Here, $V_0$ is the equilibrium volume, $B_0$ is the bulk modulus, and $B_1$ is the pressure derivative of the bulk modulus, all obtained by fitting the EOS to the Birch--Murnaghan equation. The coefficients $w_Y$ are weighting factors assigned to each parameter, with $w_{V_0}=1$, $w_{B_0}=1/20$, and $w_{B_1}=1/400$.

All ONCV pseudopotential calculations were performed with ABINIT \cite{Verstraete2025}. The $\nu$-metric verification followed the protocol of Bosoni et al. \cite{bosoni2024verify}: Fermi--Dirac smearing with a smearing temperature of $T_0=2.25$ mHa was employed, a $\Gamma$-centered uniform k-point grid with a reciprocal-space spacing finer than $\sim0.06$ \AA$^{-1}$ was used, the plane-wave cutoff energy was set to 100 Ha, and the self-consistent field (SCF) convergence threshold was $10^{-10}$ Ha/atom. The AE reference data were taken as the average of the latest cross-code AE datasets obtained by FLEUR and WIEN2k.
In addition, Tables \ref{tab:Ce_PP_nu} and \ref{tab:Er_PP_nu} include the $\nu$ values reported by Bosoni et al. \cite{bosoni2024verify} for VASP (v6.3) \cite{kresse1993ab, kresse1994ab, kresse1996efficiency, kresse1996software, kresse1999ultrasoft} using the POTPAW$\_$PBE.54 pseudopotentials and for Quantum ESPRESSO using the Standard Solid-State Pseudopotentials (SSSP) library \cite{prandini2018precision} (PBE precision, version 1.3).

\newpage

\begin{table}[h]
\caption{\label{tab:Ce_PP_nu} Ce ONCV pseudopotential performance ($\nu$ metric; X=Ce) \\ 
\textbf{Alt text}: Comparison of the Ce ONCV pseudopotential used in this work with VASP PAW and SSSP PAW pseudopotentials, based on the $\nu$ metric, across four mono-elemental cubic crystals and six cubic oxides. Our ONCV pseudopotentials have lower $\nu$ than SSSP PAW but higher than VASP PAW pseudopotentials.}
\begin{ruledtabular}
\begin{tabular*}{\textwidth}{@{\extracolsep{\fill}}cccc}
Structure & This work (ONCV) & VASP (PAW) \cite{kresse1993ab, kresse1994ab, kresse1996efficiency, kresse1996software, kresse1999ultrasoft} & SSSP (PAW) \cite{prandini2018precision} \\
\hline
SC      & 0.302 & 0.217 & 1.167 \\
BCC     & 1.041 & 0.434 & 1.098 \\
Diamond & 0.065 & 0.106 & 0.703 \\
FCC     & 0.819 & 0.232 & 0.907 \\
X$_2$O     & 0.142 & 0.088 & 0.468 \\
XO      & 0.058 & 0.056 & 0.268 \\
X$_2$O$_3$    & 0.148 & 0.113 & 0.368 \\
XO$_2$     & 0.264 & 0.121 & 0.369 \\
X$_2$O$_5$    & 0.246 & 0.144 & 0.343 \\
XO$_3$     & 0.071 & 0.185 & 0.975 \\
\hline
Avg $\nu$ & 0.316 & 0.170 & 0.667 \\
\end{tabular*}
\end{ruledtabular}
\end{table}

\newpage

\begin{table}[h]
\caption{\label{tab:Er_PP_nu} Er ONCV pseudopotential performance ($\nu$ metric, X=Er) \\ 
\textbf{Alt text}: Comparison of the Er ONCV pseudopotential used in this work with VASP PAW and SSSP PAW pseudopotentials, based on the $\nu$ metric, across four mono-elemental cubic crystals and six cubic oxides. Our ONCV pseudopotentials have lower $\nu$ than SSSP PAW but higher than VASP PAW pseudopotentials.}
\begin{ruledtabular}
\begin{tabular*}{\textwidth}{@{\extracolsep{\fill}}cccc}
Structure & This work (ONCV) & VASP (PAW) \cite{kresse1993ab, kresse1994ab, kresse1996efficiency, kresse1996software, kresse1999ultrasoft} & SSSP (PAW) \cite{prandini2018precision} \\
\hline
SC      & 1.814 & 0.904 & 4.528 \\
BCC     & 2.022 & 0.763 & 3.796 \\
Diamond & 2.850 & 2.503 & 10.349 \\
FCC     & 2.009 & 0.890 & 4.517 \\
X$_2$O     & 0.345 & 0.266 & 0.815 \\
XO      & 0.177 & 0.126 & 0.309 \\
X$_2$O$_3$    & 0.237 & 0.137 & 0.619 \\
XO$_2$     & 0.350 & 0.276 & 0.964 \\
X$_2$O$_5$    & 0.187 & 0.152 & 0.504 \\
XO$_3$     & 0.080 & 0.593 & 2.004 \\
\hline
Avg $\nu$ & 1.007 & 0.661 & 2.840 \\
\end{tabular*}
\end{ruledtabular}
\end{table}

\newpage

\section{Choice of exchange-correlation functional}

\begin{figure}[h]
    \centering
    \includegraphics[width=1\linewidth]{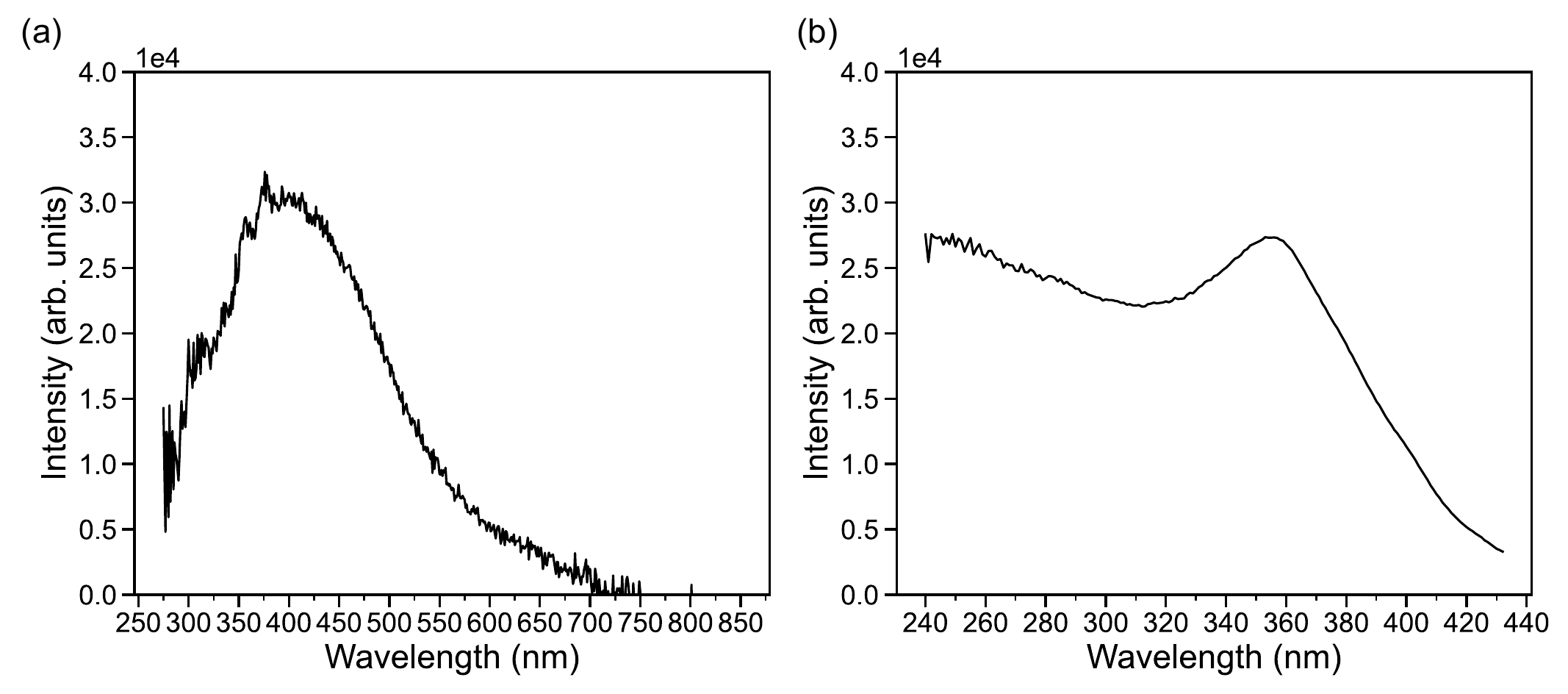}
    \caption{(a) Photoluminescence (PL; excitation at 230 nm, 5 nm bandwidth) and (b) photoluminescence excitation (PLE; collection at 450 nm) spectra of undoped single crystal $\sim$ 230 nm thick CeO$_2$ films at room temperature measured in this work. Band edge is at $\sim$360 nm (3.44 eV).\\
    \textbf{Alt text}: Photoluminescence and photoluminescence excitation spectra of undoped CeO$_2$ films plotted as intensity vs. wavelength. Both plots show the highest intensity at $\sim$360 nm.}
    \label{fig:pl_ple}
\end{figure}

Table \ref{tab:exp_lit_BG_opt} lists the experimentally measured optical bandgap of CeO$_2$. 
From Table \ref{tab:exp_lit_BG_opt}, it is apparent that the optical bandgap and dielectric constants of CeO$_2$ significantly depend on deposition parameters, substrate, and sample morphology. For example, in CeO$_2$ nanoparticles and nanocrystalline thin films, quantum confinement effects can increase the optical bandgap, causing a blue shift, but the presence of Ce$^{3+}$ ions in the sample can lead to a red shift \cite{tatar2008synthesis, patsalas2003structure}. The reported bandgap also depends on whether it was inferred from Tauc plot fits, absorption onsets, or absorption peaks, as also discussed by Castleton et al. \cite{castleton2007tuning}. All the optical bandgaps reported in Table \ref{tab:exp_lit_BG_opt} were taken from Tauc plot fits; except for  \cite{pelli2020ultrafast} and \cite{marabelli1987covalent} (who considered absorption peaks) and \cite{goubin2004experimental} (who considered absorption onset). We use the value of $\varepsilon_\infty=5.31$ as reported for bulk single crystals of CeO$_2$ \cite{mochizuki1982infrared} for the exchange fraction ($\alpha=\varepsilon_\infty^{-1}$) used in the DDH functional. As discussed in Section VII and shown in Tables \ref{tab:comp_lit_BG} and \ref{tab:piecewise_lin}, the chosen exchange fraction gives rise to a piecewise linear functional and accurate optical bandgaps (e.g., the first excitation energy computed with TDDFT@DDH is 3.49 eV, similar to the values reported for bulk powder and single crystals in Table \ref{tab:exp_lit_BG_opt} and as found in our photoluminescence and photoluminescence excitation measurements of undoped single crystal CeO$_2$ films in Figure \ref{fig:pl_ple}).   

Abbreviations and acronyms used in Tables \ref{tab:exp_lit_BG_opt} and \ref{tab:exp_lit_BG_fund}:

UV-vis DRS: ultraviolet-visible diffuse reflectance spectroscopy; EELS: electron energy loss spectroscopy; opt. refl.: optical reflectance spectroscopy; Rests. refl.: Reststrahlen reflectivity spectroscopy; SE: spectroscopic ellipsometry; UV-TAS: ultraviolet transient absorption spectroscopy; HREELS: high resolution electron energy loss spectroscopy; RF: radio-frequency sputtering; e-beam evap.: electron beam evaporation; PLD: pulsed laser deposition; co-evap.: co-evaporation; MOCVD: metal organic chemical vapor deposition; PSE-CVD: pulsed-spray evaporation chemical vapor deposition; ED: electrodeposition; PED: pulsed electron deposition; MBE: molecular beam epitaxy; PE-CVD: plasma-enhanced chemical vapor deposition; VB XPS: valence band X-ray photoelectron spectroscopy; BIS: Bremsstrahlung isochromat spectroscopy; STM: scanning tunneling microscopy; UPS: ultraviolet photoelectron spectroscopy; laser abl: laser ablation; ALD: atomic layer deposition

\newpage

\newpage

\begin{table}[h!]
\tiny
\caption{\label{tab:exp_lit_BG_opt} Experimental optical bandgap, lattice parameter, and dielectric constants of CeO$_2$ (see Section III for acronyms and abbreviations) \\
\textbf{Alt text}: Summary of several properties of CeO$_2$ samples, including single crystals and thin films, synthesized and measured using various techniques.}
\begin{ruledtabular}
\begin{tabular*}{\textwidth}{@{\extracolsep{\fill}}ccccccc}
Sample & Experiment & Bandgap (eV) & $a_0$ (\AA) & $\varepsilon_\infty$ & $\varepsilon_0$ & Notes \\
\hline

Bulk powder \cite{goubin2004experimental, zhang2006optical} & 
\shortstack{UV-vis DRS \cite{goubin2004experimental} \\ EELS \cite{goubin2004experimental} \\ UV-vis \cite{zhang2006optical}} & 
\shortstack{3.15--3.2 \cite{goubin2004experimental} \\ 3.55 \cite{zhang2006optical}}
&  & 5.52 \cite{goubin2004experimental} &  
& \shortstack{particle size \\ 1--5 $\mu$m \cite{goubin2004experimental} \\ 0.7--1 $\mu$m \cite{zhang2006optical} } \\

\hline
\shortstack{Single crystal \\ bulk \cite{marabelli1987covalent, mochizuki1982infrared} \\ thin film \cite{guo1995spectroscopic}} & 
\shortstack{Opt. refl. \cite{marabelli1987covalent} \\ Rests. refl. \cite{mochizuki1982infrared} \\ SE \cite{guo1995spectroscopic}}
& \shortstack{3.8 \cite{marabelli1987covalent} \\ 3.3--3.6 \cite{guo1995spectroscopic}} 
& & \shortstack{4.7 \cite{marabelli1987covalent} \\ 5.31 \cite{mochizuki1982infrared} \\ 6.1 \cite{guo1995spectroscopic}} 
& \shortstack{35.3 \cite{marabelli1987covalent} \\ 24.5 \cite{mochizuki1982infrared}} & \shortstack{from melt (mm$^3$) \cite{marabelli1987covalent, mochizuki1982infrared} \\ RF sputtering \cite{guo1995spectroscopic} \\ (10--100 nm thick)} \\

\hline
\shortstack{Polycrystalline\\ thin film} & 
\shortstack{UV-vis \cite{elidrissi2000structural, debnath2007optical, bueno1997optical, ruiz2013optical, hogarth1986optical, sundaram1990optical, zheng1993optical} \\ SE \cite{toro2004relationship}} &
\shortstack{3.53--3.6 \cite{elidrissi2000structural} \\ 3.6--4.11 \cite{debnath2007optical} \\ 3.12--3.16 \cite{bueno1997optical} \\ 3.33--3.66 \cite{ruiz2013optical} \\ 3.10 \cite{hogarth1986optical} \\ 3.02--3.38 \cite{sundaram1990optical} \\ 3.23 \cite{zheng1993optical} \\ 3.05 \cite{toro2004relationship}}
& 
\shortstack{5.42 \cite{elidrissi2000structural} \\ 5.50--5.54 \cite{bueno1997optical} \\ 5.3925 \cite{ruiz2013optical}} 
& \shortstack{5.06-5.76 \cite{bueno1997optical} \\ 5.38 \cite{zheng1993optical} \\ 5.22 \cite{toro2004relationship}}&  & \shortstack{spray pyrolysis \cite{elidrissi2000structural} \\ (0.5--1 $\mu$m thick; \\ 8--50 nm crystallite size) \\ e-beam evap. \cite{debnath2007optical}\\ (0.14--0.18 $\mu$m thick; \\15--36 nm crystallite size) \\ RF sputtering \cite{bueno1997optical} \\ (0.31--0.37 $\mu$m thick; \\ 68--89 nm grains) \\ PLD \cite{ruiz2013optical} \\ (0.2 $\mu$m thickness) \\ co-evap. \cite{hogarth1986optical} \\ (0.3 $\mu$m thick) \\ ox. Ce film \cite{sundaram1990optical} \\ (0.1-0.2 $\mu$m thick) \\ RF sputtering \cite{zheng1993optical}\\(0.36 $\mu$m thick) \\ MOCVD \cite{toro2004relationship} \\ (1.2--1.5 $\mu$m thick; \\ 100--130 nm grains)} \\ 

\hline
\shortstack{Nanocrystalline\\ thin film} & 
\shortstack{UV-vis \cite{ansari2010optical, jiang2009changes, zimou2021structural, acosta2019nanostructured, yang2018structural, murugan2015effect, tatar2008synthesis,orel1994optical, ozer2001optical} \\ UV-TAS \cite{pelli2020ultrafast} \\ SE \cite{barreca2003nanostructure, chiu2010optical} \\ HREELS \cite{pfau1994electronic}} &
\shortstack{3.23 \cite{ansari2010optical} \\ 2.8--3.15 \cite{jiang2009changes} \\ 3.18 \cite{zimou2021structural} \\ 2.9--3.35 \cite{acosta2019nanostructured} \\ 3.19--3.75 \cite{yang2018structural} \\ 3.45--3.53 \cite{murugan2015effect} \\ 2.58 \cite{tatar2008synthesis} \\ 3--3.6 \cite{orel1994optical} \\ 3.1 \cite{ozer2001optical} \\ 4 \cite{pelli2020ultrafast} \\2.4--2.8 \cite{barreca2003nanostructure} \\ 3.23 \cite{chiu2010optical} \\ 3.2 \cite{pfau1994electronic}} 
& 
\shortstack{5.401 \cite{zimou2021structural} \\ 5.42 \cite{tatar2008synthesis}} 
& \shortstack{4.7 \cite{barreca2003nanostructure} \\ 5.43 \cite{chiu2010optical}} &  & 
\shortstack{sol-gel \cite{ansari2010optical} \\ (3--4 nm crystallite size) \\ PSE-CVD \cite{jiang2009changes} \\ (41--334 nm thick; \\ 6--32 nm crystallite size) \\ spray pyrolysis \cite{zimou2021structural} \\ (7 nm crystallite size) \\ sol-gel \cite{acosta2019nanostructured}\\ (182--217 nm thick; \\ 4--10 nm crystallite size) \\ ED \cite{yang2018structural}\\ (5--6 nm crystalltie size) \\ RF sputtering \cite{murugan2015effect} \\ (380--590 nm thick; \\ 6--9 nm crystallite size) \\ PED \cite{tatar2008synthesis} \\ (100 nm thick; \\ 5-20 nm crystallite size) \\ sol-gel \cite{orel1994optical} \\ (120-560 nm thick) \\ sol-gel \cite{ozer2001optical} \\ (200 nm thick) \\ MBE \cite{pelli2020ultrafast} \\ (6 nm thick) \\PE-CVD \cite{barreca2003nanostructure} \\ (6--7 nm crystallite size) \\ RF sputtering \cite{chiu2010optical} \\ (12.5 nm thick) \\ e-beam evap. \cite{pfau1994electronic} \\ (10-40 nm thick)}\\ 

\hline
Nanoparticles & 
\shortstack{UV-vis DRS \cite{choudhury2015annealing, khan2014defect} \\ UV-vis \cite{wang2007remarkable, liao2001preparation}} & \shortstack{2.57--2.93 \cite{choudhury2015annealing} \\ 3.36 \cite{khan2014defect} \\ 3.13--3.66 \cite{wang2007remarkable} \\ 4.25 \cite{liao2001preparation}} & \shortstack{5.404--5.422 \cite{choudhury2015annealing} \\ 5.408 \cite{khan2014defect} \\ 5.426 \cite{wang2007remarkable}} & & & \shortstack{particle size \\ 5--14 nm \cite{choudhury2015annealing} \\ 10--100 nm \cite{khan2014defect} \\ 3.5 nm \cite{wang2007remarkable} \\ 2 nm \cite{liao2001preparation}} \\ 
\end{tabular*}
\end{ruledtabular}
\end{table}

\newpage

Table \ref{tab:exp_lit_BG_fund} lists the experimentally measured fundamental bandgap of CeO$_2$.  There are far fewer measurements of the fundamental band gap than of the optical bandgap. Valence band X-Ray photoemission spectroscopy (VB-XPS), ultra-violet photoemission spectroscopy (UPS), and scanning tunneling microscopy (STM) measurements are highly surface-sensitive: for example, surface adsorbates, surface orientation, and band-bending strongly influence measured values \cite{zhang2024environment}. The data listed in Table \ref{tab:exp_lit_BG_fund} were inferred by approximating the slope of the steepest part of the leading edge of the VBM/4f/5d bands, and thus, are highly approximate values. We also report the $\mathrm{Ce_{Ce}^{1-}}$ polaron vertical transition levels (VTLs) $\mu(-1/0)_{\mathrm{pol}}$  with respect to the VBM where available. Note that from Table 1 in the main text, the $\mathrm{Ce_{Ce}^{1-}}$ VTLs computed in this work are $\simeq$ 2.32 -- 2.81 eV with respect to the VBM (depending on whether we consider an isolated $\mathrm{Ce_{Ce}^{1-}}$ or $\mathrm{Ce_{Ce}^{1-}}$ bound to oxygen vacancy). Table \ref{tab:exp_lit_BG_fund} also lists the polaron formation energy ($E_\mathrm{self-trap}$) from ultraviolet transient absorption spectroscopy (UV-TAS), -0.4 eV.

\newpage

\begin{table}[h!]
\tiny
\caption{\label{tab:exp_lit_BG_fund} Experimental fundamental bandgaps and $\mathrm{Ce_{Ce}^{1-}}$ VTLs wrt VBM of CeO$_2$ (all energies in eV; see Section III for acronyms and abbreviations)\\
\textbf{Alt text}: Summary of fundamental bandgaps, polaron vertical transition levels, and polaron self-trapping energy of CeO$_2$ samples, measured using VB XPS-BIS, STM, UPS, and UV-TAS.}
\begin{ruledtabular}
\begin{tabular*}{\textwidth}{@{\extracolsep{\fill}}cccccc}
Experiment & \shortstack{Bandgap\\(O2p-Ce 4f)}  & \shortstack{Bandgap\\(O2p-Ce 5d)} & \shortstack{$\mathrm{Ce_{Ce}^{1-}}$ VTL \\ w.r.t. VBM} & $E_\mathrm{self-trap}$ & Notes \\
\hline
VB XPS-BIS \cite{wuilloud1984spectroscopic} & \shortstack{3.5} & \shortstack{6} & & & ox. Ce film \\
\hline
STM \cite{jerratsch2011electron} & \shortstack{4.25} & \shortstack{6} & \shortstack{1.6 (VTL)} & & \shortstack{ox. Ce film forming\\ 1 nm thick CeO2(111)} \\ 
\hline
UPS \cite{pfau1994electronic} &  &  & 2 (VTL) & & \shortstack{e-beam evap. \\ 10-40 nm thick CeO2 (111)}\\
\hline
VB XPS \cite{mullins1998electron} & & & 2 (VTL) & & \shortstack{laser abl. \\single crystal thin film CeO2 (001), (110)} \\ 
\hline
VB XPS \cite{zhang2024environment} & & & 3--3.2 (VTL) & & \shortstack{ALD \\ 20 nm thin film CeO2 (111) \\ surface carbonate species could increase VTL by 0.48 eV} \\
\hline
UV-TAS \cite{pelli2020ultrafast} & & & & -0.4 eV & \shortstack{MBE \\ 6 nm thin film} \\ 
\end{tabular*}
\end{ruledtabular}
\end{table}

\newpage

Finally, Table \ref{tab:comp_lit_BG} reports the fundamental and optical bandgaps obtained using various first-principles methods. We note here that computational studies often try to tune the Hubbard $\mathrm{U_{eff}}$ or exchange fraction $\alpha$ to match the computed Kohn-Sham gap to the optical bandgap of CeO$_2$. However, the Kohn-Sham gap, albeit approximate, should be compared to the fundamental gap, not the optical gap, as it does not account for excitonic effects, which reduce the optical gap compared to the fundamental gap \cite{sun2021pros} (the exciton binding energy in CeO$_2$ has not been reported experimentally). Based on the values reported in Table \ref{tab:comp_lit_BG} and the discussion of Section VII, we conclude that our simulation settings accurately capture the electronic structure of CeO$_2$. 

\newpage

\begin{table}[h]
\tiny
\caption{\label{tab:comp_lit_BG} Computed lattice parameter, bandgap, polaron self-trapping energy and vertical transition levels w.r.t CBM  of CeO$_2$ (all energies in eV; definitions of Eq. 5 and Eq. 6 given in Section VII) \\
\textbf{Alt text}: Summary of computed properties of CeO$_2$ with various functionals.}
\begin{ruledtabular}
\begin{tabular*}{\textwidth}{@{\extracolsep{\fill}}cccccc}
Functional & $a_0$ (\AA) & \shortstack{Bandgap\\(O2p-Ce 4f)} & \shortstack{Bandgap\\(O2p-Ce 5d)} & $E_\mathrm{self-trap}$ & \shortstack{$\mathrm{Ce_{Ce}^{1-}}$ VTL\\w.r.t. CBM} \\
\hline
LDA \cite{hohenberg1964inhomogeneous, kohn1965self} & \shortstack{5.360 \cite{hay2006theoretical} \\ 5.37 \cite{da2007hybrid}} & \shortstack{1.6 \cite{hay2006theoretical} \\ 2.0 \cite{da2007hybrid}} & \shortstack{6 \cite{hay2006theoretical} \\ 5.61 \cite{da2007hybrid}} & &  \\
\hline
PBE \cite{perdew1996generalized} & \shortstack{5.468 \cite{hay2006theoretical} \\ 5.465 \cite{du2018screened} \\ 5.47 \cite{da2007hybrid}} & \shortstack{1.7 \cite{hay2006theoretical} \\ 1.87 \cite{du2018screened} \\ 2.0 \cite{da2007hybrid}} & \shortstack{6 \cite{hay2006theoretical} \\ 5.63 \cite{du2018screened} \\ 5.64 \cite{da2007hybrid}} &  &  \\
\hline
\shortstack{PBE+U \cite{anisimov1991band, liechtenstein1995density, dudarev1998electron} \\ (U$_\mathrm{eff}$=5 eV \cite{du2018screened,sun2017disentangling}) \\ (U$_\mathrm{eff}$=4.5 eV \cite{da2007hybrid})} & \shortstack{5.500 \cite{du2018screened} \\ 5.49 \cite{da2007hybrid, sun2017disentangling}} & \shortstack{2.25 \cite{du2018screened, da2007hybrid} \\ 2.3 \cite{sun2017disentangling}} & \shortstack{5.20 \cite{du2018screened} \\ 5.3 \cite{da2007hybrid,sun2017disentangling}} & -0.54 \cite{sun2017disentangling} & 0.97 \cite{sun2017disentangling}\\
\hline
\shortstack{HSE03 \cite{heyd2003hybrid} \\ ($\alpha=0.25, \omega=0.3$ \AA$^{-1}$)} & 5.408 \cite{hay2006theoretical} & 3.3 \cite{hay2006theoretical} & 7 \cite{hay2006theoretical} &  &   \\
\hline
\shortstack{HSE06 \cite{paier2006screened} \\ ($\alpha=0.25, \omega=0.2$ \AA$^{-1}$)}\\ & \shortstack{5.394 \cite{du2018screened} \\ 5.40 \cite{da2007hybrid,sun2017disentangling}} & \shortstack{3.61 \cite{du2018screened,castleton2019benchmarking} \\ 3.5 \cite{da2007hybrid,sun2017disentangling}} & \shortstack{6.99 \cite{du2018screened} \\ 6.96 \cite{da2007hybrid} \\ 7.0 \cite{sun2017disentangling}} & \shortstack{-0.30 \cite{sun2017disentangling,castleton2019benchmarking} (Eq. 5)\\ -0.36 \cite{castleton2019benchmarking} (Eq. 6)} & 0.82 \cite{sun2017disentangling}  \\
\hline
\shortstack{HSE06 \cite{paier2006screened} \\ ($\alpha=0.15, \omega=0.2$ \AA$^{-1}$)}\\ & 5.418 \cite{du2018screened} & 2.85 \cite{du2018screened} & 6.44 \cite{du2018screened} &  &   \\
\hline
\shortstack{PBE0 \cite{adamo1999toward, ernzerhof1999assessment} \\ ($\alpha=0.25$)} & \shortstack{5.39 \cite{da2007hybrid} \\ 5.40 \cite{graciani2011comparative}} & \shortstack{4.5 \cite{da2007hybrid} \\ 4.3 \cite{graciani2011comparative} \\ 4.4 \cite{castleton2019benchmarking}} & \shortstack{7.93 \cite{da2007hybrid} \\ 8.52 \cite{graciani2011comparative}} & \shortstack{-1.45 \cite{castleton2019benchmarking} (Eq. 5) \\ -1.8 \cite{castleton2019benchmarking} (Eq. 6)} &   \\
\hline
\shortstack{G0W0@LDA+U \\ (U$_{\mathrm{eff}}$=5.4)} & & \shortstack{3.62 \cite{jiang2018revisiting} \\ 4.3 \cite{jiang2009localized}} & 6.1 \cite{jiang2009localized}   \\
\hline
\shortstack{GW0@LDA+U \\ (U$_{\mathrm{eff}}$=5.4)} & & 3.88 \cite{jiang2018revisiting} & & &  \\
\hline
G0W0@PBE & & 4.02 \cite{sun2021pros} & 6.06 \cite{sun2021pros} & &  \\
\hline
GW-BSE & & 3.88 \cite{sun2021pros} & & &  \\
\hline
TDDFT@HSE06 & & 3.76 \cite{sun2021pros} & & &  \\
\hline
TDDFT@PBE0 & & 3.80 \cite{sun2021pros} & & &  \\
\hline
PBE (this work) \cite{perdew1996generalized} & & 2.06 & 5.67 & &  \\
\hline
HSE06 (this work) \cite{paier2006screened} & & 4.26 & 6.75 & \shortstack{-0.55 (Eq. 5) \\ -0.53 (Eq. 6)} &  \\
\hline
DDH (this work) \cite{skone2014self, skone2016nonempirical} & & 4.21 & 7.03 & \shortstack{-0.37 (Eq. 5) \\ -0.62 (Eq. 6)} &  1.40 \\
\hline
TDDFT@DDH (this work) & & 3.49 & & &  \\


\end{tabular*}
\end{ruledtabular}
\end{table}

\newpage

\section{Calculation of formation energies}

We recall that the general expression for the formation energy of a defect in charge state $q$ and configuration $\mathbf{R}_{q'}$ is given by \cite{freysoldt2014first, falletta2020finite}

\begin{equation}
E_\mathrm{f}\left(q, \mathbf{R}_{q'}\right) = E\left(q, \mathbf{R}_{q'}\right) - E_\mathrm{perf} 
 + q\left(\varepsilon_\mathrm{F} + \varepsilon_\mathrm{{VBM}}\right) - \sum_i n_i \mu_i
   + E_{\mathrm{corr}}\left(q, \mathbf{R}_{q'}\right),
\label{eq:formn_E}
\end{equation}

where $E\left(q, \mathbf{R}_{q'}\right)$ is the total energy of the supercell with the defect ($\mathbf{R}_{q'}$ obtained by relaxing the atomic coordinates of the supercell with the same defect, but with charge $q'$); $E_\mathrm{perf}$ is the total energy of the pristine (perfect) supercell of the same size; $\varepsilon_\mathrm{F}$ is the Fermi level relative to the valence band maximum (VBM) $\varepsilon_\mathrm{VBM}$ of the pristine cell; $n_i$ is the number of atoms of type $i$ added to $(n_i>0)$ or removed from $(n_i<0)$ the supercell with corresponding chemical potentials $\mu_i$; and $E_{\mathrm{corr}}\left(q, \mathbf{R}_{q'}\right)$ is a correction term for the artificial electrostatic interactions arising due to the finite size of the periodic supercell. Equation \ref{eq:formn_E} allows us to compute the formation energy of a defect with charge $q$ relaxed to its equilibrium atomic configuration (required to obtain {\it thermodynamic} charge transition levels), as well as the formation energy of a defect with charge $q$ in a configuration frozen in the relaxed equilibrium atomic configuration of $q'$ (required to obtain {\it optical} charge transition levels). We use the expression proposed by Freysoldt, Neugebauer, and Van de Walle to compute $E_{\mathrm{corr}}\left(q, \mathbf{R}_{q}\right)$ \cite{freysoldt2009fully} (with $\varepsilon_0=24.5$ \cite{mochizuki1982infrared} for CeO$_2$) as implemented in sxdefectalign \cite{sxdefectalign}. The evaluation of $E_{\mathrm{corr}}\left(q, \mathbf{R}_{q'}\right)$ requires special care, as different ionic and electronic screening are involved when considering a charge $q$ in a fixed configuration corresponding to that of the charge $q'$ \cite{gake2020finite, falletta2020finite}. We compute $E_{\mathrm{corr}}\left(q, \mathbf{R}_{q'}\right)$ using the expression of Gake et al. \cite{gake2020finite} (with $\varepsilon_{\infty}=5.31$ \cite{mochizuki1982infrared} for CeO$_2$), also implemented in sxdefectalign \cite{sxdefectalign}.

\newpage

\section{Calculation of photoionization cross-sections}

The photoionization cross-section is the transition rate per unit photon flux and is computed as \cite{stoneham2001theory, razinkovas2021photoionization}
\begin{equation}
    \sigma_{\text{ph}}(\hbar\omega) = \frac{4\pi^{2}\alpha}{3 n_{\mathrm{ref}}} \hbar\omega \sum_j |\mathbf{r}_{dj}|^{2} A(\hbar\omega - E_{dj}),
    \label{eq:photoion_cs}
\end{equation}
where $\alpha$ is the fine-structure constant, $n_{\mathrm{ref}}$ is the refractive index of CeO$_2$ (2.30), and $\hbar\omega$ is the photon energy. Here we consider the photoionization of an ensemble of randomly oriented defects; $d$ denotes the initial defect state with wavefunction $\varphi_d$ and the sum runs over all final states $\varphi_j$ (i.e., the entire Ce 4f band). $\mathbf{r}_{dj}$ is the optical matrix element (computed as the transition dipole moment) between the initial defect state and final Ce 4f band state and is given by $\mathbf{r}_{dj}=\langle\varphi_d|\mathbf{r}|\varphi_j\rangle$. In principle, $\varphi_d$ and $\varphi_j$ are many-body states, but in practice, we use Kohn-Sham states \cite{razinkovas2021photoionization}. We compute $\mathbf{r}_{dj}$ as \(\mathbf{r}_{dj}=\frac{\langle\varphi_d|[\hat{H},\mathbf{r}]|\varphi_j\rangle}{\varepsilon_d - \varepsilon_j}\), where $\varepsilon_d$ ($\varepsilon_j$) is the energy of the Kohn-Sham orbital of the defect (final) state, as implemented in the WEST code \cite{govoni2015large, yu2022gpu, marcks2024quantum}. $E_{dj}$ is the energy difference between the Kohn-Sham eigenvalues of the initial and final states, rigidly shifted so that the smallest value of $E_{dj}$ corresponds to the CTL. $A(\hbar\omega - E_{dj})$ is the normalized spectral function of electron-phonon coupling, given as   
\begin{equation}
    \label{eq:spect_fn}
    A(\hbar\omega - E_{dj}) = \sum_{m} \sum_{n} P_{d}^{m}(T) \left| \langle \chi_{d}^{m} \mid \chi_{j}^{n} \rangle \right|^{2} 
    \delta \bigl( E_{dj} + E_{j}^{n} - E_{d}^{m} - \hbar\omega \bigr).
\end{equation}

We used the one-dimensional configurational coordinate diagram (1D-CCD) approach to obtain the effective phonon frequencies of the ground and excited states. The vibrational overlap function $\langle \chi_{d}^{m} \mid \chi_{j}^{n} \rangle$ between $m^{th}$ ($n^{th}$) vibrational state of defect state $\varphi_d$ (excited state $\varphi_j$) with energy $E_{d}^{m}$ ($E_{j}^{n}$) was computed using the recursion method \cite{ruhoff1994recursion}. The delta function was modeled as a Gaussian with width 25 meV. $P_{d}^{m}(T)$ is the thermal distribution function of the vibrational energy in the ground state; \(P_{d}^{m}(T) = e^{-\frac{E_{d}^m}{k_B T}}/\sum_m e^{-\frac{E_{d}^m}{k_B T}}\).

Since ionization occurs due to the excitation of the $\mathrm{Ce_{Ce}^{1-}}$ polaron (isolated or as a component of a defect complex), we use the Kohn-Sham wavefunctions and energies of the ground state configuration of the defect to capture the polaronic state of the defects when computing $\mathbf{r}_{dj}$. 
We note that we have not tested the convergence of the cross-section as a function of the supercell size, which would be a computationally prohibitive task  due to the large plane-wave cutoff (200 Ry) and use of hybrid functionals. However, we do not expect the trends to change with larger cell sizes, given the localized nature of the defects considered here. 

\newpage

\section{Localized erbium 4f states}

\begin{figure}[h]
    \centering
    \includegraphics[width=0.8\linewidth]{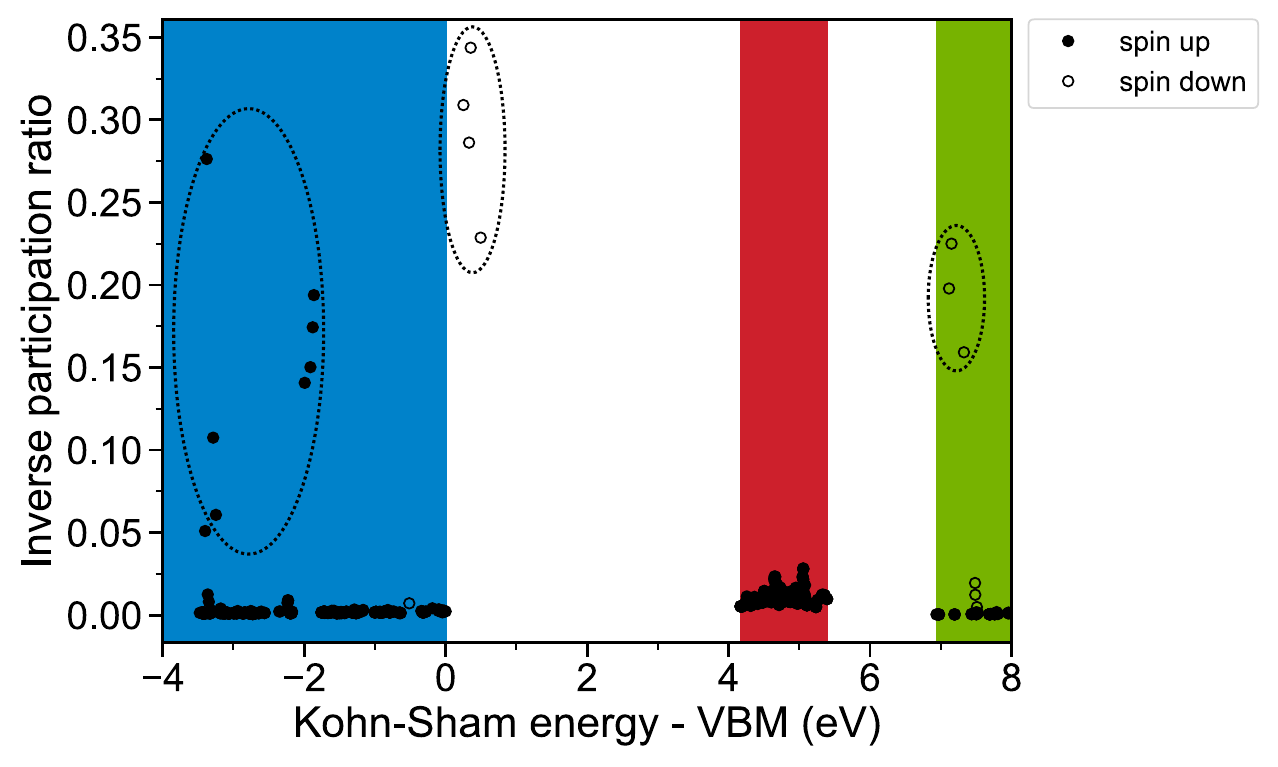}
    \caption{Inverse Participation Ratio (IPR) of Kohn-Sham orbitals in Er$^{3+}$-doped CeO$_2$. Er$^{3+}$ 4f levels are localized, as marked by the dotted ovals. Blue- valence band, red- Ce 4f band, green- Ce 5d band, VBM- valence band maximum. \\
\textbf{Alt text}: Inverse participation ratio of Kohn-Sham orbitals of Er$^{3+}$-doped CeO$_2$. Localized occupied spin up orbitals seen from $-$4 to $-$2 eV, localized occupied spin-down orbitals seen just above the VBM, and localized unoccupied spin down orbitals seen just above the start of the Ce 5d band. Ce 4f band lies between the O 2p valence band and Ce 5d band, from 4.2--5.4 eV.}
    \label{fig:loc_fac}
\end{figure}

Figure \ref{fig:loc_fac} shows the Inverse Participation Ratio (IPR) of Kohn-Sham orbitals in Er$^{3+}$-doped CeO$_2$. We see that the Er$^{3+}$ 4f levels are localized, as marked by the dotted ovals. The unoccupied spin down states occur in the Ce 5d band, not the Ce 4f band. The energy difference between the occupied and unoccupied spin down orbitals is similar to that found in Er$_2$O$_3$ using GW$_0$@LDA+U \cite{jiang2012electronic}.

\newpage

\section{Effect of the exchange-correlation functional and finite-size corrections on polaron self-trapping energy and piecewise linearity}

In CeO$_2$, an electron polaron is formed when an excess electron localizes on a Ce$^{4+}$ site, converting it to Ce$^{3+}$. The self-trapping energy of an electron polaron (also known as polaron formation energy or polaron binding energy) can be computed as 

\begin{equation}
    E_{\mathrm{self\text{-}trap}} = E\left(-1, \mathbf{R}_{-1}\right) - E\left(-1, \mathbf{R}_{\mathrm{perf}}\right)
    \label{eq:self_trap_no_corr}
\end{equation}

or 

\begin{equation}
E_{\mathrm{self\text{-}trap}} = E\left(-1, \mathbf{R}_{-1}\right) - E_\mathrm{perf} - \varepsilon_\mathrm{CBM} + E_{\mathrm{corr}}\left(-1, \mathbf{R}_{-1}\right),
\label{eq:self_trap_corr}
\end{equation}

Here, $E\left(-1, \mathbf{R}_{-1}\right)$ is the total energy of the supercell with an excess electron that localizes to form the polaron (and $E_{\mathrm{corr}}\left(-1, \mathbf{R}_{-1}\right)$ is its corresponding finite-size correction); $E\left(-1, \mathbf{R}_{\mathrm{perf}}\right)$ is the total energy of the pristine supercell with an excess delocalized electron. Note that $\varepsilon_\mathrm{CBM}$ is the minimum of the Ce 4f band. A negative $E_{\mathrm{self\text{-}trap}}$ indicates that the polaron is more stable than a delocalized electron in the conduction band. Rearranging the terms in Equation 1 of the main text (see calculation of CTL) and Equation \ref{eq:self_trap_corr} above results in $\varepsilon(0/-1)_{\mathrm{pol}}=-E_{\mathrm{self-trap}}$.

$E_{\mathrm{self\text{-}trap}}$ computed using Equation \ref{eq:self_trap_no_corr} gives $-0.37$ eV, similar to the value ($-0.30$ eV) obtained using HSE06 by Sun et al. \cite{sun2017disentangling} and Castleton \cite{castleton2019benchmarking}. $E_{\mathrm{self\text{-}trap}}$ computed using Equation \ref{eq:self_trap_corr} gives $-0.62$ eV. In principle, with an exact exchange-correlation functional, piecewise linearity would be satisfied \cite{janak1978proof, harbola1999relationship, falletta2025equivalence} (i.e., \(E\left(-1, \mathbf{R}_{\mathrm{perf}}\right)-E_\mathrm{perf}\) would be equal to $\varepsilon_\mathrm{CBM}$ of the pristine supercell), and thus Equations \ref{eq:self_trap_no_corr} and \ref{eq:self_trap_corr} should give the same value (assuming $E_{\mathrm{corr}}\left(-1, \mathbf{R}_{-1}\right)$ is accounted for in Equation \ref{eq:self_trap_no_corr}). The discrepancy between the two values above is due to the missing finite-size corrections in Equation \ref{eq:self_trap_no_corr}. 

Equation \ref{eq:self_trap_no_corr} assumes that the finite-size corrections for the supercell with the delocalized electron is equal to that of the cell with the localized electron polaron (i.e., $E_{\mathrm{corr}}\left(-1, \mathbf{R}_{\mathrm{perf}}\right)=E_{\mathrm{corr}}\left(-1, \mathbf{R}_{-1}\right)$), and therefore does not include any corrections. However, as reported in multiple recent studies \cite{gake2020finite, falletta2020finite}, this is an incorrect assumption, as the electronic and ionic polarizations for the delocalized system are different from those of the localized system, resulting in different finite-size corrections. Thus, Equation \ref{eq:self_trap_no_corr} should be re-written as

\begin{equation}
    E_{\mathrm{self\text{-}trap}} = E\left(-1, \mathbf{R}_{-1}\right) - E\left(-1, \mathbf{R}_{\mathrm{perf}}\right) + E_{\mathrm{corr}}\left(-1, \mathbf{R}_{-1}\right) - E_{\mathrm{corr}}\left(-1, \mathbf{R}_{\mathrm{perf}}\right)
    \label{eq:self_trap_corr_2}
\end{equation}

$E_{\mathrm{corr}}\left(-1, \mathbf{R}_{-1}\right)$ is calculated using the method of Freysoldt, Neugebauer, and Van de Walle \cite{freysoldt2009fully} using $\varepsilon_0=24.5$ \cite{mochizuki1982infrared}, and $E_{\mathrm{corr}}\left(-1, \mathbf{R}_{\mathrm{perf}}\right)$ is calculated using the method of Gake et al. \cite{gake2020finite} using $\varepsilon_{\infty}=5.31$ \cite{mochizuki1982infrared}.  In particular, in the calculation of $E_{\mathrm{corr}}\left(-1, \mathbf{R}_{\mathrm{perf}}\right)$, the excess electron is only electronically screened. Since the charge is delocalized, only the point-charge correction term needs to be included, accounting for the interaction of the background neutralizing charge with the defect potential. We used the implementation in the sxdefectalign package to obtain the finite-size corrections.

As seen in Table \ref{tab:piecewise_lin}, $E_{\mathrm{self\text{-}trap}}$ calculated using Equation \ref{eq:self_trap_corr_2} is nearly equal to that calculated using Equation \ref{eq:self_trap_corr}. This agreement results from the fact that including $E_{\mathrm{corr}}\left(-1, \mathbf{R}_{\mathrm{perf}}\right)$ restores piecewise linearity, as shown in Table \ref{tab:piecewise_lin}, where \(E\left(-1, \mathbf{R}_{\mathrm{perf}}\right)-E_\mathrm{perf}+E_{\mathrm{corr}}\left(-1, \mathbf{R}_{\mathrm{perf}}\right)\) is nearly equal to the $\varepsilon_\mathrm{CBM}$ of the pristine supercell. 

Similarly, if piecewise linearity is satisfied, the Kohn-Sham defect level of the polaron should remain constant while changing its occupation (from $-$1 to 0 in the case of electron polaron), with the supercell geometry fixed to that with the polaronic distortion. As seen in Table \ref{tab:piecewise_lin}, including finite-size corrections improves the agreement between the Kohn-Sham levels, with the difference between $\varepsilon_\mathrm{pol}(-1)$ and $\varepsilon_\mathrm{pol}(0)$ reducing from 0.44 eV to 0.26 eV. The correction to $\varepsilon_\mathrm{pol}(-1)$ and $\varepsilon_\mathrm{pol}(0)$ were calculated as $\frac{-2E_{\mathrm{corr}}\left(-1, \mathbf{R}_{-1}\right)}{-1}$ and $\frac{2E_{\mathrm{corr}}\left(0, \mathbf{R}_{-1}\right)}{-1}$, respectively, as described by Falletta et al. \cite{falletta2025equivalence}.

\begin{table}[h]
\tiny
\caption{\label{tab:piecewise_lin} $E_{\mathrm{self\text{-}trap}}$ values from various formalisms and demonstration of the effect of finite-size corrections on piecewise linearity using the CBM and polaron Kohn-Sham level \\
\textbf{Alt text}: Effect of finite-size corrections on polaron self-trapping energy and piecewise linearity. Polaron self-trapping energies computed using different formalisms give the same value only upon including finite-size corrections. Similarly, piecewise linearity in the CBM and polaron Kohn-Sham levels is observed only after including the corrections.}
\begin{ruledtabular}
\begin{tabular*}{\textwidth}{@{\extracolsep{\fill}}cccccc}
Quantity & Value & Quantity & Value & Quantity & Value  \\
\hline
$E_{\mathrm{self\text{-}trap, Eq. 4}}$           & -0.37 & $\varepsilon_\mathrm{CBM}$           & 14.55 & $\varepsilon_\mathrm{pol}(-1)$ & 13.52 \\
$E_{\mathrm{self\text{-}trap, Eq. 5}}$           & -0.62 & \(E\left(-1, \mathbf{R}_{\mathrm{perf}}\right)-E_\mathrm{perf}\) & 14.24 & $\varepsilon_\mathrm{pol}(0)$ & 13.96 \\
$E_{\mathrm{self\text{-}trap,Eq. 6}}$ & -0.66 & \(E\left(-1, \mathbf{R}_{\mathrm{perf}}\right)-E_\mathrm{perf}+E_{\mathrm{corr}}\left(-1, \mathbf{R}_{\mathrm{perf}}\right)\) & 14.60 & $\varepsilon_\mathrm{pol}(-1)$ (w/ corr) & 13.66 \\
                                            &       &                                        &   & $\varepsilon_\mathrm{pol}(0)$ (w/ corr) & 13.40 \\
\end{tabular*}
\end{ruledtabular}
\end{table}

We note here that the deviation from piecewise linearity is larger for the polaron Kohn-Sham level than for the CBM. This could be due to strain-related finite-size effects in our  2~$\times$~2~$\times$~2 supercell, which are not accounted for by the electrostatics-based correction schemes employed in this work. For instance, Castleton \cite{castleton2019benchmarking} showed that the polaron formation energy computed using LDA+U and a 96-atom supercell can differ by $0.06-0.13$ eV from the thermodynamic limit, even with finite-size corrections. 

Finally, we note that we do not see any discrepancies when using Equation \ref{eq:self_trap_no_corr} vs. Equation \ref{eq:self_trap_corr} with the HSE06 functional (see Table \ref{tab:comp_lit_BG}) - possibly due to the screening incorporated in HSE06, which likely (and spuriously) corrects for the finite-size effects when computing $E_{\mathrm{self\text{-}trap}}$ using Equation \ref{eq:self_trap_no_corr} (note that in the benchmarking study by Castleton \cite{castleton2019benchmarking}, $E_{\mathrm{self\text{-}trap}}$ computed using Equation \ref{eq:self_trap_no_corr} vs. Equation \ref{eq:self_trap_corr} with the PBE0 exchange-correlation functional also shows a sizeable difference, similar to that found with the DDH functional in this work).  

\newpage

\section{Configuration coordinate diagrams for the calculation of defect ionization}

\begin{figure}[h!]
    \centering
    \includegraphics[width=1\linewidth]{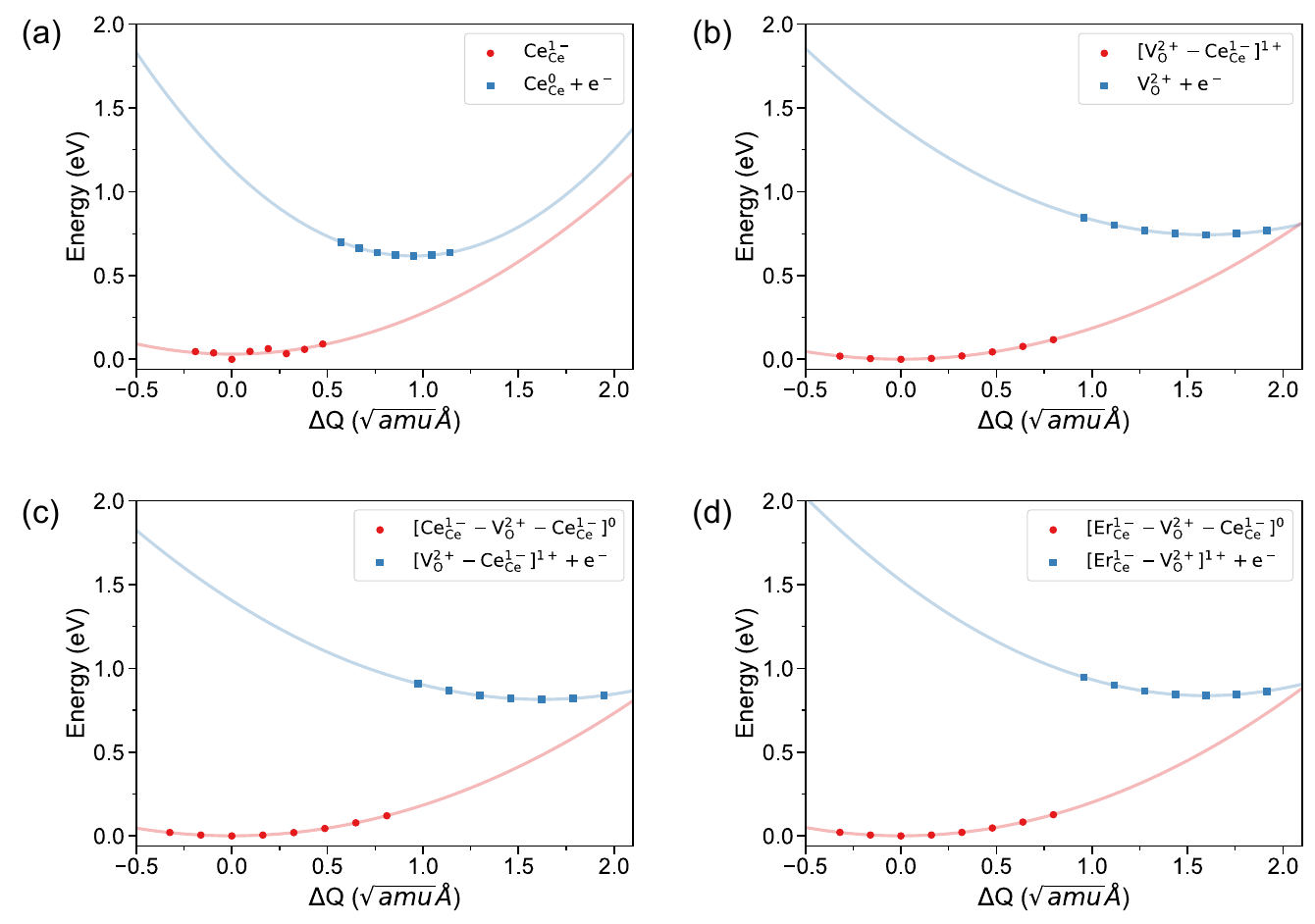}
    \caption{One-dimensional configuration coordinate diagrams (1D-CCD) of the various defect ionizations considered in this study. Only points used to fit the parabolas are shown. $\Delta Q$ is the mass-weighted displacement, calculated as $\Delta Q = \left( \sum_{\alpha = 1}^{N} \sum_{i = x, y, z} M_{\alpha} \Delta R_{\alpha i}^{2} \right)^{1/2}$, where $M_{\alpha}$ = mass of atom $\alpha$ and $\Delta R_{\alpha i}$ = displacement of atom $\alpha$ between the excited state and ground state equilibrium structures in the $i^{\mathrm{th}}$ direction).\\
    \textbf{Alt text}: 1D-CCD plots for the ionization of the various defects considered in this work (see Table 1 in main text). The points near the relaxed ground- and excited-state structures are used to fit the potential energy surfaces.}
    \label{fig:ccd}
\end{figure}

\newpage 

\begin{table}[h]
\caption{\label{tab:vibronic}
Ground- and excited-state effective vibrational frequencies
($\hbar\omega_{gs}$ and $\hbar\omega_{es}$), and mass-weighted 
displacements ($\Delta Q$) along the effective phonon mode for the defect transitions above.
Frequencies are given in meV and $\Delta Q$ in amu$^{1/2}$\,\AA. Defects are described using the expression $\mathrm{A_B}^q$, denoting atom A substituting site B with relative charge $q$. $\mathrm{V_O}$ denotes oxygen vacancy.\\
\textbf{Alt text}: Ground- and excited-state effective phonon frequencies and mass-weighted 
displacements ($\Delta Q$) along the effective phonon mode of the various defect ionizations considered in this work. The complexes have similar frequencies ($\sim$40 meV in the ground state, $\sim$45 meV in the excited state), with $\Delta Q$ $\sim$ 1.60. The frequencies of the isolated polaron are slightly higher and the $\Delta Q$ is lower.
}
\begin{ruledtabular}
\begin{tabular}{l r r r}
Transition & $\hbar\omega_{gs}$ & $\hbar\omega_{es}$ & $\Delta Q$ \\
\hline
Ce$_{\mathrm{Ce}}^{1-} \rightarrow$
Ce$_{\mathrm{Ce}}^{0} + e^{-}$
& 45.28 & 69.42 & 0.95 \\

V$_{\mathrm O}^{1+} \rightarrow$
V$_{\mathrm O}^{2+} + e^{-}$
& 39.31 & 46.00 & 1.60 \\

V$_{\mathrm O}^{0} \rightarrow$
V$_{\mathrm O}^{1+} + e^{-}$
& 39.10 & 43.29 & 1.62 \\

[Er$_{\mathrm{Ce}}$--V$_{\mathrm O}$]$^{0}
\rightarrow$
[Er$_{\mathrm{Ce}}$--V$_{\mathrm O}$]$^{1+} + e^{-}$
& 40.85 & 47.48 & 1.60 \\
\end{tabular}
\end{ruledtabular}
\end{table}

\newpage

\section{Photoionization cross-sections without vibrational broadening and with vibrational broadening at 4 K and 300 K}
\begin{figure}[h]
    \centering
    \includegraphics[width=0.8\linewidth]{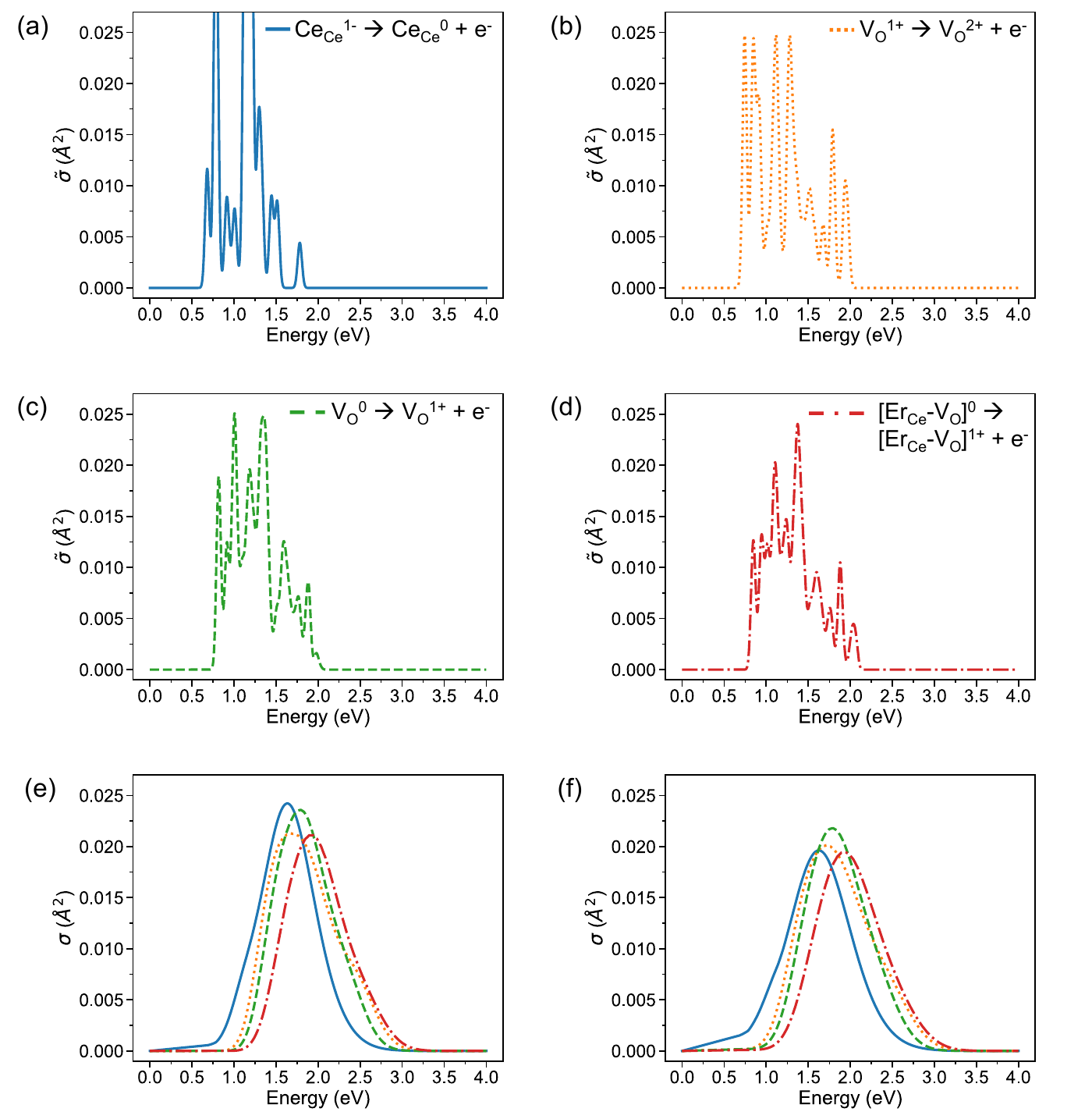}
    \caption{(a)-(d) Photoionization cross-sections without vibrational broadening (obtained by replacing $A(\hbar\omega - E_{dj})$ in Equation \ref{eq:photoion_cs} with $\delta(\hbar\omega - E_{dj})$) showing finite cross-sections at 0.8 eV. (e) Photoionization cross-sections with vibrational broadening at 4 K and (f) 300 K. Cross-section at 4 K is similar to that at 300 K, except that the latter is slightly broadened and red-shifted. Defects are described using the expression $\mathrm{A_B}^q$, denoting atom A substituting site B with relative charge $q$. $\mathrm{V_O}$ denotes oxygen vacancy.\\
    \textbf{Alt text}: Photoionization cross-sections of the defect transitions in this work. Onset seen at the CTL when vibrational broadening is not included. Onset is blue-shfted when vibrational broadening is included. Cross-sections also become smooth when vibrational broadening is included.}
    \label{fig:photoion_cs_temp}
\end{figure}

\newpage

\section{Photocurrent measurements}
\begin{figure}[h]
    \centering
    \includegraphics[width=1\linewidth]{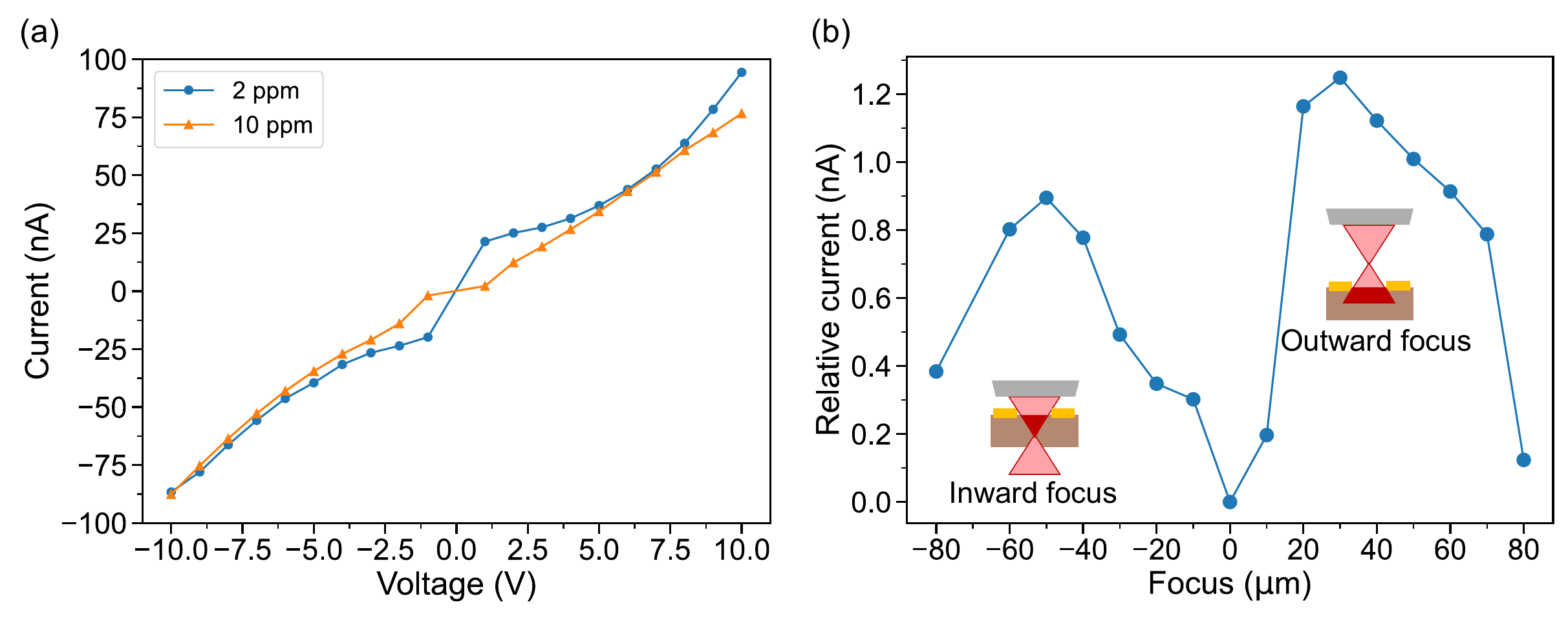}
    \caption{(a) Typical I-V curve at 0.1 V/s scan rate, showing largely similar behavior of the 2 ppm and 10 ppm doped samples. (b) Current varies with depth of laser focus; positive and negative values correspond to focus out of the sample and into the sample, respectively. Sample surface is at 0 $\mu$m. Relative current is defined as $I_{\mathrm{relative}}=I-I_{\mathrm{focus=0}}$, where $I$ and $I_{\mathrm{focus=0}}$ are the total current and current measured with the laser focus on the surface, respectively. Insets in (b) adapted from Heremans et al. \cite{heremans2009generation}. Solid lines between data points are a guide to the eye. \\
    \textbf{Alt text}: Figure on the left plots the I-V curve of 2 and 10 ppm doped samples, showing nearly equal currents at a given voltage. Figure on the right plots current as a function of laser focus depth, showing higher current when focus is out of the sample.}
    \label{fig:photocurrent_tests}
\end{figure}

Figure \ref{fig:photocurrent_tests}a shows a standard I-V measurement on 2 and 10 ppm doped Er$^{3+}$-CeO$_2$ samples, demonstrating that their carrier density is largely similar. 

For Er$^{3+}$-CeO$_2$ sample synthesis, see Ref. \cite{grant2024optical}.

For photocurrent device fabrication, samples were cleaned with isopropanol and deionized (DI) water, then primed with hexamethyldisilazane (HMDS) prior to spin-coating a bi-layer resist stack (LOR 3A and Microposit 1805, prebaked at 180$^\circ$C and 115$^\circ$C respectively).  Device patterns were written using laser lithography (Heidelberg MLA) and developed in CD-26 for 60 seconds.  The developed stacks were then coated with Ti/Au (10/150nm) via e-beam evaporation.  Lift-off was performed to reveal the patterned devices by soaking in 80$^\circ$C NMP for 1 hour, followed by mild sonication (10 sec). The devices were cleaned with acetone and isopropanol, then mounted to carrier circuit boards and wire-bonded.

To confirm that the photocurrent is dominated by carriers from the CeO$_2$ bulk (and not the surface), we measured the photocurrent with varying focus under a fixed 1.2 mW laser power in the 10 ppm sample, shown in Figure \ref{fig:photocurrent_tests}b. The slightly higher photocurrent when the laser is focused out of the sample compared to when it is focused into the sample indicates that the photocurrent is proportional to the illuminated volume, confirming the dominant bulk effect. 

We note here that the photocurrent is likely carried by electrons. The consumption of oxygen from the CeO$_2$ lattice due to the formation of the interfacial CeO$_\mathrm{x}$-SiO$_\mathrm{y}$ layer leaves behind $\mathrm{V_O^{2+}}$ compensated by $\mathrm{Ce_{Ce}^{1-}}$. Under illumination, $\mathrm{Ce_{Ce}^{1-}}$ ionizes to form free electrons in the conduction band, and due to the applied voltage, the electrons in the conduction band drift to the positive electrode and $\mathrm{V_O^{2+}}$ drifts to the negative electrode. Electrons are then resupplied from the outer circuit back into CeO$_2$. In principle, Ce interstitials (e.g., $\mathrm{Ce_i^{4+}}$) could also form (also compensated by $\mathrm{Ce_{Ce}^{1-}}$), but their concentration is likely negligible compared to $\mathrm{V_O^{2+}}$ \cite{keating2012analysis}. Finally, for holes to contribute to photocurrent, there would have to be acceptor states (such as those from cerium vacancies $\mathrm{V_{Ce}^{4-}}$ or oxygen interstitials $\mathrm{O_i^{2-}}$) with charge transition levels $\leq$ 0.8 eV above the valence band. However, these defects are unlikely to form under our experimental conditions and their concentrations are likely negligible compared to oxygen vacancies \cite{keating2012analysis}. 

\newpage

\newpage

\bibliography{si}